\documentstyle[12pt,aaspp4]{article}

\tighten
\slugcomment{Submitted for publication in 
{\it The Astrophysical Journal}}

\begin{document}


\def\gap{\;_\sim^>\;}
\def\lap{\;_\sim^<\;}

\def\pmb#1{\setbox0=\hbox{#1}%
  \kern0.00em\copy0\kern-\wd0
  \kern0.03em\copy0\kern-\wd0
  \kern0.00em\raise.04em\copy0\kern-\wd0
  \kern0.03em\raise.04em\copy0\kern-\wd0\box0 }

\def\pp{\parshape=2 -0.25truein 6.75truein 0.5truein 6truein}

\def\ref #1;#2;#3;#4;#5{\par\pp #1 #2, #3, #4, #5}
\def\book #1;#2;#3{\par\pp #1 #2, #3}
\def\rep #1;#2;#3{\par\pp #1 #2, #3}

\def\undertext#1{$\underline{\smash{\hbox{#1}}}$}
\def\simlt{\lower.5ex\hbox{$\; \buildrel < \over \sim \;$}}
\def\simgt{\lower.5ex\hbox{$\; \buildrel > \over \sim \;$}}

\def\etal{{et~al.}}
\def\noi{\noindent}
\def\bs{\bigskip}
\def\ms{\medskip}
\def\ss{\smallskip}
\renewcommand{\deg}{$^{\circ}$}
\newcommand{\um}{$\mu$m}
\newcommand{\uk}{$\mu$K}
\newcommand{\qrms}{$Q_{\rm rms-PS}$}
\newcommand{\n}{$n$}
\newcommand{\cdmr}{${\bf c}_{\rm DMR}$}
\newcommand{\xrms}{$\otimes_{RMS}$}
\newcommand{\gt}{$>$}
\newcommand{\lt}{$<$}
\newcommand{\ldl}{$< \delta <$}
\newcommand{\sgm}{$(\delta M/M)[8 h^{-1}\ {\rm Mpc}]$}

\begin{flushright}
MIT-CTP-2548, KUNS 1399 \hskip 0.5cm August 1996
\end{flushright}
\bs
\title{${\bf COBE}$-DMR-NORMALIZED OPEN CDM COSMOGONIES}

\author{
Krzysztof M. G\'orski\altaffilmark{1,2,3}, 
Bharat Ratra\altaffilmark{4,5},
Rados{\l}aw Stompor\altaffilmark{6,7}, 
Naoshi Sugiyama\altaffilmark{8},
and
A.J. Banday\altaffilmark{9}
}

\noindent
\altaffiltext{1}{Hughes STX Corporation, Code 685, LASP, NASA/GSFC,
Greenbelt, MD 20771.}
\altaffiltext{2}{Theoretical Astrophysics Center, Juliane Maries Vej 30,
2100 Copenhagen \O, Denmark.}
\altaffiltext{3}{Warsaw University Observatory, Aleje Ujazdowskie 4, 
00-478 Warszawa, Poland.}
\altaffiltext{4}{Center for Theoretical Physics, Massachusetts
Institute of Technology, Cambridge, MA 02139.}
\altaffiltext{5}{Department of Physics, Kansas State University,
Manhattan, KS 66505.}
\altaffiltext{6}{Department of Physics, University of Oxford, Keble
Road, Oxford, OX1 3RH, UK.}
\altaffiltext{7}{Copernicus Astronomical Center, Bartycka 18, 00-716
Warszawa, Poland.}
\altaffiltext{8}{Department of Physics, Kyoto University,
Kitashirakawa-Oiwakecho, Sakyo-ku, Kyoto 606, Japan.}
\altaffiltext{9}{Max Planck Institut f\"ur Astrophysik,
Karl-Schwarzschild-Strasse 1, 85740 Garching bei M\"unchen, Germany.}
\bigskip
\bigskip
\bigskip

\begin{abstract}

Cut-sky orthogonal mode analyses of the $COBE$-DMR 53 and 90 GHz sky 
maps are used to determine the normalization of a variety of open 
cosmogonical models based on the cold dark matter scenario. To
constrain the allowed cosmological-parameter range for these open
cosmogonies, the predictions of the DMR-normalized models are compared
to various observational measures of cosmography and large-scale 
structure, viz.: the age of the universe; small-scale dynamical estimates
of the clustered-mass density parameter $\Omega_0$; constraints on the 
Hubble parameter $h$, the X-ray cluster baryonic-mass fraction $\Omega_B/
\Omega_0$, and the matter power spectrum shape parameter; estimates of the 
mass perturbation amplitude; and constraints on the large-scale peculiar 
velocity field. 

The open-bubble inflation model (Ratra \&\ Peebles 1994;
Bucher, Goldhaber, \&\ Turok 1995; Yamamoto, Sasaki, \&\ Tanaka 1995) is
consistent with current determinations of the 95\% confidence level (c.l.)
range of these observational constraints. More specifically, for a range
of $h$, the model is reasonably consistent with recent high-redshift 
estimates of the deuterium abundance which suggest $\Omega_B h^2 \sim 0.007$,
provided $\Omega_0 \sim 0.35$; recent high-redshift estimates of the 
deuterium abundance which suggest $\Omega_B h^2 \sim 0.02$ favour $\Omega_0
\sim 0.5$, while the old nucleosynthesis value $\Omega_B h^2 = 0.0125$ requires
$\Omega_0 \sim 0.4$.

Small shifts in the inferred $COBE$-DMR normalization amplitudes due to: 
(1) the small differences between the galactic- and ecliptic-coordinate 
sky maps, (2) the inclusion or exclusion of the quadrupole moment in the 
analysis, (3) the faint high-latitude Galactic emission treatment, and, 
(4) the dependence of the theoretical cosmic microwave background anisotropy 
angular spectral shape on the value of $h$ and $\Omega_B$, are explicitly 
quantified.

The DMR data alone do not possess sufficient discriminative power to prefer 
any values for $\Omega_0$, $h$, or $\Omega_B$ at the 95\% c.l. for the models 
considered. At a lower c.l., and when the quadrupole moment is included in 
the analysis, the DMR data are most consistent with either $\Omega_0 \lap 0.1$ 
or $\Omega_0 \sim 0.7$ (depending on the model considered). However, when the 
quadrupole moment is excluded from the analysis, the DMR data are most 
consistent with  $\Omega_0 \sim 0.35-0.5$ in all open models considered (with 
$0.1 \leq \Omega_0 \leq 1$), including the open-bubble inflation model. 
Earlier claims (Yamamoto \&\ Bunn 1996; Bunn \&\ White 1996) that the DMR data 
require a 95\% c.l. lower bound on $\Omega_0$ ($\sim 0.3$) are not supported 
by our (complete) analysis of the four-year data: the DMR data alone cannot be 
used to meaningfully constrain $\Omega_0$.

\end{abstract}


\section{INTRODUCTION}

Quantum-mechanical fluctuations during an early epoch of inflation provide a 
plausible mechanism to generate the energy-density perturbations responsible
for observed cosmological structure. While it has been known for quite some 
time that inflation is consistent with open spatial hypersurfaces (Gott 1982; 
Guth \&\ Weinberg 1983), attention was initially focussed on models in 
which there are a very large number of $e$-foldings during inflation, 
resulting in almost exactly flat spatial hypersurfaces for the observable 
part of the present universe (Guth 1981; also see Kazanas 1980; Sato 1981a,b).
This was, perhaps, inevitable because of strong theoretical prejudice towards 
flat spatial hypersurfaces and their resulting simplicity. However, to get a 
very large number of $e$-foldings during inflation it seems necessary that the 
inflation model have a small dimensionless parameter (J. R. Gott, private 
communication 1994; Banks et al. 1995), which would require an explanation.
 
Attempts to reconcile these ``favoured" flat spatial hypersurfaces with 
observational measures of a low value for the clustered-mass density 
parameter $\Omega_0$ have concentrated on models in which one postulates the 
presence of a cosmological constant $\Lambda$ (Peebles 1984). In the simplest 
flat-$\Lambda$ model one assumes a scale-invariant (Harrison 1970; Peebles \&\ 
Yu 1970; Zel'dovich 1972) primordial power spectrum for gaussian adiabatic 
energy-density perturbations. Such a spectrum is generated by 
quantum-mechanical fluctuations during an early epoch of inflation in a 
spatially-flat model, provided that the inflaton potential is reasonably flat 
(Fischler, Ratra, \&\ Susskind 1985, and references therein)\footnote{
In inflation models the small observed cosmic microwave background
(CMB) anisotropy could be the consequence of the small ratio of the inflation
epoch mass scale to the Planck mass (Ratra 1991, and references therein; also
see Banks et al. 1995).}.
It has been demonstrated that these models are indeed 
consistent with current observational constraints (e.g., Stompor, G\'orski, 
\&\ Banday 1995; Ostriker \&\ Steinhardt 1995; Ratra \&\ Sugiyama 1995; 
Liddle et al. 1996b; Ganga, Ratra, \&\ Sugiyama 1996b, hereafter GRS).

An alternative, more popular of late, is to accept that the spatial 
hypersurfaces are not flat. In this case, the radius of curvature for the 
open spatial sections introduces a new length scale (in addition to the 
Hubble length), which requires a generalization of the usual flat-space 
scale-invariant spectrum (Ratra \&\ Peebles 1994, hereafter RP94). Such a 
spectrum is generated by quantum-mechanical fluctuations during an 
epoch of inflation in an open-bubble model (RP94; Ratra \&\ Peebles 1995, 
hereafter RP95; Bucher et al. 1995, hereafter BGT; Lyth \&\ Woszczyna 1995; 
Yamamoto et al. 1995, hereafter YST), provided that the inflaton potential 
inside the bubble is reasonably flat. Such gaussian adiabatic open-bubble 
inflation models have also been shown to be consistent with current 
observational constraints (RP94; Kamionkowski et al. 1994; G\'orski et al. 1995,
hereafter GRSB; Liddle et al. 1996a, hereafter LLRV; Ratra et al. 1995; GRS).

Inflation theory by itself is unable to predict the normalization amplitude 
for the energy-density perturbations. Currently, the least controversial and 
most robust method for the normalization of a cosmological model is to fix 
the amplitude of the model-predicted large-scale CMB spatial anisotropy by 
comparing it to the observed CMB anisotropy discovered by the $COBE$-DMR 
experiment (Smoot et al. 1992).

Previously, specific open cold dark matter (CDM) models have been examined
in light of the $COBE$-DMR two-year results (Bennett et al. 1994). GRSB 
investigated the CMB anisotropy angular spectra predicted by the open-bubble
inflation model (RP94), and compared large-scale structure predictions of this
DMR-normalized model to observational data.\footnote{
Ratra et al. (1995) and GRS subsequently extended the analysis to smaller 
scales, comparing detailed CMB anisotropy predictions to observational data.}
Cay\'on et al. (1996) performed a related analysis for the open model with
a flat-space scale-invariant spectrum (Wilson 1983, hereafter W83), and 
Yamamoto \&\ Bunn (1996, hereafter YB) examined the effect of additional
sources of quantum fluctuations (BGT; YST) in the open-bubble inflation model.

In this paper, we study the observational predictions for a number of open CDM
models. In particular, we employ the power spectrum estimation technique 
devised by G\'orski (1994) for incomplete sky coverage to normalize the open 
models using the $COBE$-DMR four-year data (Bennett \etal\ 1996). In $\S 2$ 
we provide an overview of open-bubble inflation cosmogonies. In $\S 3$ we 
detail the various DMR data sets used in the analyses here, discuss the 
various open models we consider, and present the DMR estimate of the CMB rms 
quadrupole anisotropy amplitude $Q_{\rm rms-PS}$ as a function of $\Omega_0$
for these open models. In $\S 4$ we detail the computation of several 
cosmographic and large-scale structure statistics for the DMR-normalized open 
models. These statistics are confronted by various current observational 
constraints in $\S 5$. Our results are summarized in $\S 6$.

\section{OPEN-BUBBLE INFLATION MODELS}

The simplest open inflation model is that in which a single open-inflation 
bubble nucleates in a (possibly) spatially-flat, inflating spacetime (Gott 
1982; Guth \&\ Weinberg 1983). In this model, the first epoch of inflation 
smooths away any preexisting spatial inhomogeneities, while simultaneously 
generating quantum-mechanical zero-point fluctuations. Then, in a tunnelling 
event, an open-inflation bubble nucleates, and for a small enough nucleation 
probability the observable universe lies inside a single open-inflation 
bubble. Fluctuations of relevance to the late-time universe can be generated 
via three different quantum mechanical mechanisms: (1) they can be generated 
in the first epoch of inflation; (2) they can be generated during the 
tunnelling event (thus resulting in a slightly inhomogeneous initial 
hypersurface inside the bubble, or a slightly non-spherical bubble); and (3) 
they can be generated inside the bubble. The tunneling amplitude is largest 
for the most symmetrical solution (and deviations from symmetry lead to an 
exponential suppression), so it has usually been assumed that the nucleation 
process 
(mechanism [2]) does not lead to the generation of significant inhomogeneities. 
Quantum-mechanical fluctuations generated during evolution inside the bubble 
(RP95) are significant. Assuming that the energy-density difference between 
the two epochs of inflation is negligible (and so the bubble wall is not 
significant), one may estimate the contribution to the perturbation spectrum 
after bubble nucleation from quantum-mechanical fluctuations during the 
first epoch of inflation (BGT; YST). As discussed by Bucher \&\ Turok (1995, 
hereafter BT) (also see YST; YB), the observable predictions of these simple 
open-bubble inflation models are almost completely insensitive to the details 
of the first epoch of inflation, for the observationally-viable range of 
$\Omega_0$. This is because the fluctuations generated during this epoch 
affect only the smallest wavenumber part of the energy-density perturbation 
power spectrum, which cannot contribute significantly to observable quantities 
because of the spatial curvature length ``cutoff" in an open universe (e.g., 
W83; Kamionkowski \&\ Spergel 1994; RP95). Inclusion of such fluctuations in 
the calculations alter the predictions for the present value of the rms linear 
mass fluctuations averaged over an $8 h^{-1}$ Mpc sphere, \sgm, by 
$\sim 0.1-0.2\%$ (which is comparable to our computational accuracy). 

Besides the open-bubble inflation model spectra, a
variety of alternatives have also been considered. Predictions for the 
usual flat-space scale-invariant spectrum in an open model have been 
examined (W83; Abbott \&\ Schaefer 1986; Gouda, Sugiyama, \&\ Sasaki 1991;
Sugiyama \&\ Gouda 1992; Kamionkowski \&\ Spergel 1994; Sugiyama \&\ Silk 1994;
Cay\'on et al. 1996). The possibility that the standard formulation of quantum
mechanics is incorrect in an open universe, and that allowance must be made 
for non-square-integrable basis functions has been investigated 
(Lyth \&\ Woszczyna 1995), and other spectra have also been considered (e.g., 
W83; Abbott \&\ Schaefer 1986; Kamionkowski \&\ Spergel 1994). These spectra, 
being inconsistent with either standard quantum mechanics or the length scale 
set by spatial curvature, are of historical interest. 

More recently, the open-bubble inflation scenario has been further elaborated 
on. YST have considered a very specific model for the nucleation of the 
open bubble in a spatially-flat de Sitter spacetime, and demonstrated a 
possible additional contribution from a non-square-integrable basis function
which depends on the form of the potential, and on the assumed form of the 
quantum state prior to bubble nucleation\footnote{
If the length scale set by the mass determined by the curvature of the 
inflaton potential in the first epoch of inflation is significantly smaller
than the Hubble length, as is expected in reasonable particle physics models,
there is no non-square-integrable basis function in the second epoch of 
inflation (YST).}.
However, since the non-square-integrable basis function contributes only on 
the very largest scales, the spatial curvature ``cutoff" in an open universe 
makes almost all of the model predictions insensitive to this basis function, 
for the observationally-viable range of $\Omega_0$ (YST; YB). For example, at 
$\Omega_0 \sim 0.4-0.5$ its effect is to change \sgm\ by 
$\sim 0.8-1\%$\footnote{
Hence it seems that there is as yet 
no need to speculate about the quantum state prior to bubble nucleation. 
However, more recently it has been suggested that in certain two field 
models (Linde \&\ Mezhlumian 1995) the contribution of this 
non-square-integrable basis function might be enhanced by the ratio of the 
energy densities before and after bubble nucleation, and it has been suggested 
that if this ratio is large it would be a problem for these two field models 
(Sasaki \&\ Tanaka 1996). However, this depends sensitively on the speculative
properties of the pre-nucleation model and vacuum state.}.

An additional possible effect determined for the specific model of an
open-inflation bubble nucleating in a spatially-flat de Sitter spacetime
is that fluctuations of the bubble wall behave like a non-square-integrable 
basis function (Hamazaki et al. 1996; Garriga 1996; Garc\'{\i}a-Bellido 1996;
Yamamoto, Sasaki, \&\ Tanaka 1996). While there are models in which these
bubble-wall fluctuations are completely insignificant (Garriga 1996; Yamamoto
et al. 1996), there is as yet no computation that accounts for both the 
bubble-wall fluctuations as well as those generated during the evolution 
inside the bubble (which are always present), so it is not yet known if
bubble-wall fluctuations can give rise to an observationally significant 
effect. Finally, again in this very specific model, the effects of
a finite bubble size at nucleation seem to alter the zero bubble size 
predictions only by a very small amount (Yamamoto et al. 1996; Cohn 1996).

While there is no guarantee that there is a spatially-flat de Sitter spacetime 
prior to bubble nucleation, these computations do illustrate the important 
point that the spatial curvature length ``cutoff" in an open universe (e.g., 
RP95) does seem to ensure that what happens prior to bubble nucleation does not 
significantly affect the observable predictions for observationally-viable 
single-field open-bubble inflation models. It is indeed reassuring that 
accounting only for the quantum mechanical fluctuations generated during the 
evolution inside the bubble (RP94) seems to be essentially all that is 
required to make observational predictions for the single-field open-bubble 
inflation models. That is, the observational predictions of the open-bubble 
inflation scenario seem to be as robust as those for the spatially-flat 
inflation scenario. 

\section{CMB ANISOTROPY NORMALIZATION PROCEDURE}

\subsection{Data Selection and Power Spectrum Inference}

In this paper, we utilize the DMR four-year 53 and 90 GHz sky maps in both 
galactic and ecliptic coordinates. We thus quantify explicitly the expected 
small shifts in the inferred normalization amplitudes due to the small 
differences between the galactic- and ecliptic-coordinate maps. The maps are 
coadded using inverse-noise-variance weights derived in each coordinate 
system. The least sensitive 31 GHz maps have been omitted from the analysis, 
since their contribution is minimal under such a weighting scheme.

The dominant source of emission in the DMR maps is due to the Galactic plane. 
We are unable to model this contribution to the sky temperature to sufficient 
accuracy to enable its subtraction, thus we excise all pixels where the 
Galactic-plane signal dominates the CMB. The geometry of the cut has been 
determined by using the DIRBE 140 $\mu$m map as a tracer of the strongest 
emission, as described completely in Banday \etal\ (1996a). All pixels with 
Galactic latitude $|b|\ <$ 20\deg$\,$ are removed, together with regions 
towards Scorpius-Ophiucus and Taurus-Orion. There are 3881 surviving pixels in 
galactic coordinates and 3890 in ecliptic. This extended (four-year data) 
Galactic plane cut has provided the biggest impact on the analysis of the DMR 
data (see G\'orski et al. 1996, hereafter G96).

The extent to which residual high-latitude Galactic emission can modify our 
results has been quantified in two ways. Since the spatial morphology of 
Galactic synchrotron, free-free and dust emission seems to be
well described by a steeply falling power spectrum ($\sim\ \ell^{-3}$ -- 
Kogut \etal\ 1996a, G96), the cosmological signal is predominantly compromised
on the largest angular scales. As a simple test of
Galactic contamination, we perform all computations
both including and excluding the observed sky quadrupole.
A more detailed approach (G96) notes that a large fraction of the Galactic 
signal can be accounted for by using the DIRBE 140 $\mu$m sky map 
(Reach \etal\ 1995) as a template for free-free and 
dust emission, and the 408 MHz all-sky radio survey 
(Haslam \etal\ 1981) to describe synchrotron emission.
A correlation analysis yields coupling coefficients
for the two templates at each of the DMR frequencies.
We have repeated our model analysis after 
correcting the coadded sky maps by the Galactic
templates scaled by the coefficients derived in G96. In particular, we adopt 
those values derived under the assumption that the CMB anisotropy is 
well-described by an $n$ = 1 power law model with normalization amplitude
\qrms\ $\sim$ 18 $\mu$K\footnote{
A more self-consistent analysis would simultaneously compute the \qrms\ and 
coupling coefficient amplitudes. In fact, we have investigated this for a 
sub-sample of the models considered here in which we varied $\Omega_0$ but 
fixed $h$ and $\Omega_B$. No statistically significant changes were found in 
the derived values of either \qrms\ or the coupling coefficients.}.
One might make criticisms of either technique: excluding information from an
analysis, in this case the quadrupole components, can obviously weaken any
conclusions simply because statistical uncertainties will grow; at the same
time, it is not clear whether the Galactic corrections applied are completely
adequate. We believe that, given these uncertainties, our analysis is the most 
complete and conservative one that is possible.

The power spectrum analysis technique developed by G\'orski (1994) is 
implemented. Orthogonal basis functions for the Fourier decomposition of the 
sky maps are constructed which specifically include both pixelization effects 
and the Galactic cut. (These are linear combinations of the usual spherical 
harmonics with multipole $\ell \leq 30$.) The functions are coordinate system 
dependent. A likelihood analysis is then performed as described in G\'orski 
\etal\ (1994).

\subsection{Theoretical Spectra of Anisotropy}

We consider four open model energy-density perturbation power spectra: 
(1) the open-bubble inflation model spectrum, accounting only for fluctuations 
that are generated during the evolution inside the bubble (RP94); (2) the 
open-bubble inflation model spectrum, now also accounting for the fluctuations
generated in the first epoch of inflation (BGT; YST); (3) 
the open-bubble inflation model spectrum, now also accounting for both the 
usual fluctuations generated in the first epoch of inflation and 
a contribution from a non-square-integrable basis function (YST); and, (4) an 
open model with a flat-space scale-invariant spectrum (W83). In all cases we 
have ignored the possibility of tilt or primordial gravity waves, since it is 
unlikely that they can have a significant effect in viable open models.

With the eigenvalue of the spatial scalar Laplacian being $-(k^2 + 1)$, where
$k\, (0 < k < \infty)$ is the radial coordinate spatial wavenumber, the 
gauge-invariant fractional energy-density perturbation power spectrum of type 
(1) above is
$$
  P(k) = A{(4 + k^2)^2 \over k(1 + k^2)} T^2 (k) ,
  \eqno(1)
$$
where $T(k)$ is the transfer function and $A$ is the normalization 
amplitude\footnote{
In the literature, the primordial part of this open-bubble inflation model 
spectrum is occasionally called the ``conformal" spectrum or the 
``scale-invariant" spectrum. These names are misleading: the open de Sitter
spacetime inside the bubble is not conformal to spatially-flat Minkowski
spacetime (more precisely, it is conformal to the upper ``Milne wedge" of
Minkowski spacetime), which is why the primordial part of the spectrum of 
eq. (1) is manifestly non-scale-invariant. This spectrum is, however, the 
``natural" generalization of the flat-space scale-invariant spectrum to the 
open model, and it is the open-bubble inflation model spectrum accounting 
only for those fluctuations generated during the evolution inside the bubble.
Note that Bunn \&\ White (1996, hereafter BW, eq. [30]) generalize the 
primordial part of the spectrum of eq. (1) by multiplying it with 
$(k^2 + 1)^{(n-1)/2}$. As yet, only the specific $n = 1$ generalized spectrum 
(i.e., eq. [1]) is known to be a prediction of an open-bubble inflation model 
and therefore consistent with the presence of spatial curvature. It is 
premature to draw conclusions about open cosmogony on the basis of the $n \neq
1$ version of the spectrum considered by BW.}. 
In the simplest example, perturbations generated in the 
first epoch of inflation introduce an additional multiplicative factor,
${\rm coth}(\pi k)$, on the right hand side of eq. (1). For a discussion of 
the effects of the non-square-integrable basis function see YST and YB. 
The energy-density power spectrum of type (4) above is
$$
  P(k) = A k T^2 (k) , 
  \eqno(2)
$$
and in this case one can also consider, e.g., $P(k) \propto \sqrt{1 + k^2}$ 
(W83), but 
because of the spatial curvature ``cutoff" in an open model the predictions 
are essentially indistinguishable\footnote{
It should be noted that such open model spectra are ``unnatural" --- they
do not account properly for the additional length scale set by the radius 
of space curvature in an open universe. We include the case of eq. (2) here
both for historical reasons and to provide a \lq\lq strawman'' 
to compare to the open-bubble inflation model.}.
At small $k$ the asymptotic expressions are $P(k) \propto k^{-1}$ (type 1), 
$P(k) \propto k^{-2}$ (type 2), and $P(k) \propto k$ (type 4). 

Conventionally, the CMB fractional temperature perturbation, $\delta T/T$, is 
expressed as a function of angular position, $(\theta, \phi)$, on the sky
via the spherical harmonic decomposition,
$$
  {\delta T \over T}(\theta , \phi) = \sum_{\ell=2}^\infty
       \sum_{m=-\ell}^\ell a_{\ell m} Y_{\ell m}(\theta , \phi) .
  \eqno(3)
$$
The CMB spatial anisotropy in a gaussian model can then be characterized by 
the angular perturbation spectrum $C_\ell$, defined in terms of the ensemble 
average, 
$$
   \langle a_{\ell m} a_{\ell^\prime m^\prime}{}^* \rangle = 
      C_\ell \delta_{\ell\ell^\prime} \delta_{mm^\prime} .
   \eqno(4)
$$

The $C_\ell$'s used here were computed using two independent Boltzmann 
transfer codes developed by NS (e.g., Sugiyama 1995) and RS (e.g., Stompor 
1994). Some illustrative comparisons are shown in Fig. 1. We emphasize that 
the excellent agreement between the $C_\ell$'s computed using the two codes is 
mostly a reflection of the currently achievable numerical accuracy. 
Currently, the major likely additional, unaccounted for, source of uncertainty
is that due to the uncertainty in the modelling of various physical effects. 
The computations here assume a standard recombination thermal history, 
and ignore the possibility of early reionization. The simplest open models 
(with the least possible number of free parameters) have yet to be ruled out 
by observational data (GRSB; Ratra et al. 1995; GRS; this paper), so there is 
insufficient motivation to expand the model-parameter space by including the 
effect of early reionization, tilt or gravity waves\footnote{
Note that the geometrical effect in an open universe moves the effects of
early reionization on the CMB anisotropy to a smaller angular scale. As a
result $Q_{\rm rms-PS}$ values determined from the DMR data here (assuming
no early reionization) are unlikely to be very significantly affected by
early reionization. However, since structure forms earlier in an open model,
other effects of early reionization might be more significant in an open model.
While it is possible to heuristically account for such effects, an accurate 
quantitative estimate must await a better understanding of structure 
formation.}.

For the $P(k)$ of types (1), (2), and (4) above, we have evaluated the CMB 
anisotropy angular spectra for a range of $\Omega_0$ spanning the 
interval between 0.1 and 1.0, for a variety of values of $h$ (the Hubble 
parameter $H_0 = 100h \, {\rm km} \, {\rm s}^{-1} {\rm Mpc}^{-1}$) and the 
baryonic-mass density parameter $\Omega_B$. The values of $h$ were selected 
to cover the lower part of the range of ages consistent with current 
requirements ($t_0 \simeq$  10.5 Gyr, 12 Gyr, or 13.5 Gyr, with $h$ as a
function of $\Omega_0$ computed accordingly; see, for example, Jimenez 
et al. 1996; Chaboyer et al. 1996). The values of $\Omega_B$ were chosen 
to be consistent with current standard nucleosynthesis requirements
($\Omega_B h^2=$  0.0055, 0.0125, or 0.0205; e.g., Copi, Schramm, \&\ 
Turner 1995; Sarkar 1996). To render the problem tractable, $C_\ell$'s 
were determined for the central values of $t_0$ and $\Omega_B h^2$, and 
for the two combinations of these parameters which most perturb the 
$C_\ell$'s from those computed at the central values (i.e., for the 
smallest $t_0$ we used the smallest $\Omega_B h^2$, and for the largest $t_0$ 
we used the largest $\Omega_B h^2$). Specific parameter values are given in 
columns (1) and (2) of Tables 1--6,
and representative anisotropy spectra can be seen in Figs. 2 and 3. We 
therefore improve on our earlier analysis of the DMR two-year data (GRSB) by 
considering a suitably broader range in the ($\Omega_B$, $h$) parameter space.

The CMB anisotropy spectra for $P(k)$ of type (3) above were computed for a 
range of $\Omega_0$ spanning the interval between 0.1 and 0.9, for $h = 0.6$ 
and $\Omega_B = 0.035$. Specific parameter values are given in columns (1) and 
(2) of Table 7, and these spectra are shown in Fig. 4. In Fig. 5 we compare
the various spectra considered here.

The differences in the low-$\ell$ shapes of the $C_\ell$'s in the various 
models (Figs. 2--5) are a consequence of three effects: (1) the shape of the 
energy-density perturbation power spectrum at low wavenumber; (2) the 
exponential suppression at the spatial curvature scale in an open model; and 
(3) the interplay between the ``usual" (fiducial CDM) Sachs-Wolfe term and 
the ``integrated" Sachs-Wolfe (hereafter SW) term in the expression for the 
CMB spatial anisotropy. The relative importance 
of these effects is determined by the value of $\Omega_0$, and leads to the 
non-monotonic behaviour of the large-scale $C_\ell$'s as a function of 
$\Omega_0$ seen in Figs. 2--5. More precisely, the contributions to the CMB 
anisotropy angular spectrum from the ``usual" and ``integrated" SW 
terms have a different $\ell$-dependence as well as a relative amplitude that 
is both $\Omega_0$ and $P(k)$ dependent.  

On very large angular scales (small $\ell$'s), the dominant contribution to the
``usual" SW term comes from a higher redshift (when the length scales
are smaller) than does the dominant contribution to the ``integrated" 
SW term (Hu \&\ Sugiyama 1994, 1995). As a result, in an open model on
very large angular scales, the ``usual" SW term is cut off more 
sharply by the spatial curvature length scale than is the ``integrated"
SW term (Hu \&\ Sugiyama 1994), i.e., on very large angular scales
in an open model the ``usual" SW term has a larger (positive) 
effective index $n$ than the ``integrated" SW term. On slightly 
smaller angular scales the ``integrated" SW term is damped (i.e., 
it has a negative effective index $n$) while the ``usual" SW term 
plateaus (Hu \&\ Sugiyama 1994). As a consequence, going from the largest
to slightly smaller angular scales, the ``usual" term rises steeply and then
flattens, while the ``integrated" term rises less steeply and then drops
(i.e., it has a peak). The change in shape, as a function of $\ell$, of these
two terms is both $\Omega_0$ and $P(k)$ dependent. These are the two dominant 
effects at $\ell \lap 15-20$; at higher $\ell$ other effects come into play.

More specifically, for $\Omega_0 > 0.8$ the curvature length scale cutoff
and the precise large-scale form of the $P(k)$ considered here are 
relatively unimportant --- the CMB anisotropy angular spectrum is quite
similar to that for $\Omega_0 = 1$, and the dominant contribution is the 
``usual" SW term. For a
$P(k)$ that does not diverge at low wavenumber, as with the flat-space 
scale-invariant spectrum in an open model, for $\Omega_0 \lap 0.8$ the 
exponential ``cutoff" at the spatial curvature length dominates, and
the lowest-$\ell$ $C_\ell$'s are suppressed (Figs. 3 and 5). For this $P(k)$,
as $\Omega_0$ is reduced, the ``usual" term continues to be important on the 
largest angular scales down to $\Omega_0 \sim 0.4-0.5$. As $\Omega_0$ is 
reduced below $\sim 0.4-0.5$ the ``integrated" term starts to dominate on the
largest angular scales, and as $\Omega_0$ is further reduced the ``integrated"
term also starts to dominate on smaller angular scales. From Fig. 3(a) one
will notice that the ``integrated" SW term ``peak" first makes an 
appearance at $\Omega_0 = 0.4$ --- the central line in the plot at
$\ell (\ell + 1) C_\ell\ +\ {\rm offset}\ \sim 3$ --- and that as $\Omega_0$
is further reduced (in descending order along the curves shown) the 
``integrated" term ``peak" moves to smaller angular scales. The $\Omega_0 \sim
0.4$ case is where the ``integrated" term peaks at $\ell \sim 2-3$, and the
damping of this term on smaller angular scales ($\ell \gap 5$) is compensated
for by the steep rise of the ``usual" SW term --- the two terms
are of roughly equal magnitude at $\ell \sim 10$ --- and these effects 
result in the almost exactly scale-invariant spectrum at $\Omega_0 \sim 0.4$
(this case is more scale-invariant than fiducial CDM). A discussion of some of
these features of the CMB anisotropy angular spectrum in the flat-space 
scale-invariant spectrum open model is given in Cay\'on et al. (1996).

Open-bubble inflation models have a $P(k)$ that diverges at low 
wavenumber (RP95; note that no physical quantity diverges), 
and this increases the low-$\ell$ $C_\ell$'s (Figs. 2 and 5) relative to those 
of the flat-space scale-invariant spectrum open model (Figs. 3 and 5). The 
$C_\ell$'s for low $\Omega_0$ models increase more than the higher $\Omega_0$ 
ones, since, for a fixed wavenumber-dependence of $P(k)$, the divergence is 
more prominent at lower $\Omega_0$ (RP94). The 
non-square-integrable basis function (YST) contributes even more power on 
large angular scales, and  so, at low-$\ell$, the $C_\ell$'s of Fig. 4 are 
slightly larger than those of Fig. 2 (also see Fig. 5). Again, spectra at 
lower values of $\Omega_0$ are more significantly influenced.

As is clear from Figs. 2 and 5, in an open-bubble inflation model, 
quantum-mechanical zero-point fluctuations generated in the first epoch of 
inflation scarcely affect the $C_\ell$'s, although at the very lowest values of 
$\Omega_0$ the very lowest order $C_\ell$ coefficients are slightly modified. 
The effect is concentrated in this region of the parameter space since the 
fluctuations in the first inflation epoch only contribute to, and increase, 
the lowest wavenumber part of $P(k)$.
In simple open-bubble inflation models, the precise value of this small 
effect is dependent on the model assumed for the first epoch of 
inflation (BT). Since the DMR data is most sensitive to 
multipole moments with $l \sim$ 8--10, one expects the effect at $l \sim$ 2--3 
to be almost completely negligible (BT; also see YST; YB). 

Figs. 3--5 show that both the flat-space scale-invariant spectrum open model, 
and the contribution from the non-square-integrable mode, do lead to 
significantly different $C_\ell$'s (compared to those of Fig. 2).

\subsection{Results of $Q_{\rm rms-PS}$ fitting}

The results of the DMR likelihood analyses are summarized in Figs. 6--21
and Tables 1--7 and 13.

Two representative sets of likelihood functions $L(Q_{\rm rms-PS}, \Omega_0)$
are shown in Figs. 6 and 7. Figure 6 shows those derived from the 
ecliptic-frame sky maps, ignoring the correction for faint high-latitude
foreground Galactic emission, and excluding the quadrupole moment from the
analysis. Figure 7 shows the likelihood functions derived from the 
galactic-frame sky maps, accounting for the faint high-latitude foreground 
Galactic emission correction, and including the quadrupole moment in the 
analysis. Together, these two data sets span the maximum range of normalizations
inferred from our analysis (the former providing the highest,
and the latter the lowest \qrms).

Tables 1--7 give the $Q_{\rm rms-PS}$ central values and 1-$\sigma$ and 
2-$\sigma$ ranges for spectra of type (1), (3), and (4) above,
computed from the appropriate posterior probability
density distribution function assuming a uniform prior.
Each line in Tables 1--7 lists these values at a given $\Omega_0$ for the 
8 possible combinations of: (1) galactic- or ecliptic-coordinate map;
(2) faint high-latitude Galactic foreground emission correction accounted for or
ignored; and, (3) quadrupole included ($\ell_{\rm min} = 2$) or excluded
($\ell_{\rm min} = 3$)\footnote{
BW have recently considered the DMR four-year data in the context of the 
open-bubble inflation model accounting only for the fluctuations generated 
during the evolution inside the bubble (RP94). However, they use an analytic
approximation to the CMB anisotropy angular spectra, only consider
the ecliptic-frame maps, ignore the correction for faint high-latitude 
Galactic foreground emission, and choose not to examine the consequences of 
exclusion of the quadrupole from the analysis. They also do not seem to
have examined the effect on the derived $Q_{\rm rms-PS}$ value of varying 
cosmological parameters like $\Omega_B$. Since they do not quote derived 
$Q_{\rm rms-PS}$ values for this model we are not able to compare to their 
results.}.
The corresponding ridge lines of maximum likelihood
$Q_{\rm rms-PS}$ value as a function of $\Omega_0$ are shown in Figs. 8--10
for some of the cosmological-parameter values considered here.

Although we have computed these values for spectra of type (2) above (i.e., 
those accounting for perturbations generated in the first epoch 
of inflation) we record only a subset of them in column (4) of Table 13. 
These should be compared to columns (2) and (6) of Table 13, which show the 
maximal 2-$\sigma$ $Q_{\rm rms-PS}$ range for spectra of types (1) and (3).
While the differences in $Q_{\rm rms-PS}$ between spectra (1) and (2) [cols. 
(2) and (4) of Table 13] are not totally insignificant, more importantly the 
differences between the \sgm\ values for the three spectra [cols. (3), (5), 
and (7) of Table 13] are observationally insignificant.

The entries in Tables 1--6 illustrate the shift in the inferred normalization 
amplitudes due to changes in $h$ and $\Omega_B$. These shifts are larger for 
models with a larger $\Omega_0$, since these models have CMB anisotropy spectra 
that rise somewhat more rapidly towards large $\ell$, so in these cases 
the DMR data is sensitive to somewhat smaller angular scales where the effects
of varying $h$ and $\Omega_B h^2$ are more prominent. Figure 11 shows the 
effects that varying $t_0$ and $\Omega_B h^2$ have on some of the ridge 
lines of maximum likelihood $Q_{\rm rms-PS}$ as a function of $\Omega_0$,
and Fig. 13 illustrates the effects on some of the conditional (fixed
$\Omega_0$ slice) likelihood densities for $Q_{\rm rms-PS}.$ On the whole, 
for the CMB anisotropy spectra considered here, shifts in $h$ and $\Omega_B 
h^2$ have only a small effect on the inferred normalization amplitude.

The normalization amplitude is somewhat more sensitive to the differences 
between the galactic- and ecliptic-coordinate sky maps, to the foreground
high-latitude Galactic emission treatment, and to the inclusion or exclusion of
the $\ell = 2$ moment. See Figs. 8--10. For the purpose of normalizing models, 
we choose for our 2-$\sigma$ c.l. bounds values from the likelihood fits that 
span the maximal range in the \qrms\ normalizations. Specifically, 
for the lower 2-$\sigma$ bound we adopt the value determined from the 
analysis of the galactic-coordinate maps accounting for the high-latitude 
Galactic emission correction and including the $\ell = 2$ moment in the 
analysis, and for the upper 2-$\sigma$ value that determined from the analysis
of the ecliptic-coordinate maps ignoring the Galactic emission correction and 
excluding the $\ell = 2$ moment from the analysis.
These values are recorded in columns (5) and (8) of Tables 9--12, and 
columns (2), (4), and (6) of Table 13\footnote{
Since different grids (in $Q_{\rm rms-PS}$) were used in the likelihood 
analyses of the various model spectra, and different interpolation methods
were used in the determination of the $Q_{\rm rms-PS}$ values, there are 
small (but insignificant) differences in the quoted $Q_{\rm rms-PS}$ values
for some identical models in these tables.}.

Figure 12 compares the ridge lines of maximum likelihood $Q_{\rm rms-PS}$
value, as a function of $\Omega_0$, for the four different CMB anisotropy
angular spectra considered here, and Fig. 14 compares some of the conditional
(fixed $\Omega_0$ slice) likelihood densities for $Q_{\rm rms-PS}$ for 
these four CMB anisotropy angular spectra.

Approximate fitting formulae may be derived to describe the above two extreme 
2-$\sigma$ limits. For the open-bubble inflation model (RP94; BGT; YST), not 
including a contribution from a non-square-integrable basis function, we have
$$
 Q_{\rm rms-PS} (\Omega_0)/\mu{\rm K} \simeq 19^{+3.50}_{-3.25}
      + \left(4.95^{+1.1}_{-1.2} - \Omega_0\right)
        {\rm sin} [ 2\pi \{ 1 + 0.25 (1.1 - \Omega_0)\} \{\Omega_0 - 0.05\}],
 \eqno(5)
$$
which is good to better than $\sim 5\%$ for all values of $\Omega_0$
(and to better than $\sim 2\%$ over the observationally-viable range of 
$0.3 \lap \Omega \lap 0.6$). For those models including a contribution 
from the non-square-integrable basis function (YST), we have
$$
 Q_{\rm rms-PS} (\Omega_0)/\mu{\rm K} \simeq 21^{+3.7}_{-4.0} 
      + \left( 5.55^{+1.1}_{-1.1} - \Omega_0\right)
        {\rm cos} [ 1.25\pi \{1 + 0.25 (1.1 - \Omega_0)\} \{\Omega_0 - 0.05\}],
 \eqno(6)
$$
mostly good to better than $\sim 2\%$. The flat-space scale-invariant 
spectrum open model fitting formula is
$$
 Q_{\rm rms-PS} (\Omega_0)/\mu{\rm K} \simeq 15^{+2.95}_{-2.50}
      + \left( 3.25^{+0.6}_{-0.8}\right)
        {\rm sin} [ 2\pi (1 + 0.25\Omega_0) (\Omega_0 + 0.05) - 1.25 ] ,
 \eqno(7)
$$
generally good to better than $\sim 4\%$, except near $\Omega_0 \sim 0.1$
and $\Omega_0 \sim 1$ where the deviations are larger. Further details
about these fitting formulae may be found in Stompor (1996). 

The approximate fitting formulae (5)--(7) provide a convenient, portable
normalization of the open models. It is important, however, to note that they
have been derived using the $Q_{\rm rms-PS}$ values determined for a given 
$h$ and $\Omega_B$, and hence do not account for the additional uncertainty
(which could be as large as $\sim 2\%$) due to allowed variations in these 
parameters. We emphasize that in our analysis here we make 
use of the actual $Q_{\rm rms-PS}$ values derived from the likelihood 
analyses, not these fitting formulae.

Figures 15 and 16 show projected likelihood densities for 
$\Omega_0$, for some of the models and DMR data sets considered here.
Note that the general features of the projected likelihood densities 
for the open-bubble inflation model only accounting for the fluctuations
generated during the evolution inside the bubble (spectrum [1] above),
are consistent with those derived from the DMR two-year data (GRSB, Fig. 3).
However, since we only compute down to $\Omega_0 = 0.1$ here, only the 
rise to the prominent peak at very low $\Omega_0$ (GRSB) is seen.
BW show in the middle left-hand panel of their Fig. 11 (presumably) the 
projected likelihood density for $\Omega_0$ for the same open-bubble inflation 
model, the general features of which are consistent with those derived here. 

Figures 17--21 show marginal likelihood densities for
$\Omega_0$, for some of the models and DMR data sets considered here. 
For the open-bubble inflation model accounting only for the fluctuations 
generated during the evolution inside the bubble (RP94), the DMR two-year 
data galactic-frame (quadrupole moment excluded and included) marginal 
likelihoods are shown in Fig. 3 of GRSB, and are in general concord with 
those shown in Fig. 17 here (although, again, only the rise to the prominent 
low-$\Omega_0$ peak is seen here). Note that now, especially for the 
quadrupole excluded case, the peaks and troughs are more prominent (although 
still not greatly statistically significant). Furthermore, comparing the 
solid line of Fig. 17(b) here to the heavy dotted line of Fig. 3 of GRSB, one 
notices that the intermediate $\Omega_0$ peak is now at $\Omega_0 \lap 0.4$, 
instead of at $\Omega_0 \sim 0.5$ for the DMR two-year data. (Since BW 
chose not to compute for the case when the quadrupole moment is excluded from 
the analysis, they presumably did not notice the peak at $\Omega_0 \sim 
0.35-0.45$ in the marginalized likelihood density for the open-bubble 
inflation model --- see Fig. 17.) 

For the open-bubble inflation model now also accounting for both the 
fluctuations generated in the first spatially-flat epoch of inflation
(BGT; YST), and those from the non-square-integrable basis function (YST), 
the DMR two-year data ecliptic-frame quadrupole-included marginal likelihood 
(shown as the solid line in Fig. 3 of YB) is in general agreement with the 
dot-dashed line of Fig. 19(a). However, YB did not compute for the case where 
the quadrupole moment was excluded from the analysis and so did not find the 
peak at $\Omega_0 \sim 0.4-0.45$ in Fig. 19. 

Given the shapes of the marginal likelihoods in Figs. 17--21, 
it is not at all clear if it is meaningful to derive limits on 
$\Omega_0$ without making use of other (prior) information. As an example,
it is not at all clear what to use for the integration range in $\Omega_0$. 
Focussing on Fig. 21(a) (which is similar to the other quadrupole excluded 
cases), the only conclusion seems to be that $\Omega_0 \sim 0.4$ is the value 
most consistent with the DMR data (at least amongst those models with 
$0.1 \leq \Omega_0 \leq 1$ --- some of the models have another peak at 
$\Omega_0 < 0.1$, GRSB). However, when the quadrupole moment is included in 
the analysis (as in Fig. 21b), the open-bubble inflation model peaks are at 
$\Omega_0 \sim 0.7$ (at least in the range $0.1 \leq \Omega_0 \leq 1$, GRSB),
while the flat-space scale-invariant spectrum open model peak is at 
$\Omega_0 \lap 0.1$. At the 95\% c.l. no value of $\Omega_0$ over the range 
considered, 0.1--1, is excluded. (The YB and BW claims of a lower limit on 
$\Omega_0$ from the DMR data alone are, at the very least, premature.) 

\section{COMPUTATION OF LARGE-SCALE STRUCTURE STATISTICS}

The $P(k)$ (e.g., 
eqs. [1] and [2]) were determined from a numerical integration of the 
linear perturbation theory equations of motion. As before, the computations were
performed with two independent numerical codes. For some of the model-parameter
values considered here the results of the two computations were compared
and found to be in excellent agreement. Illustrative examples of the
comparisons are shown in Fig. 22. Again, we emphasize that the excellent 
agreement is mostly a reflection of the currently available numerical 
accuracy, and the most likely additional, unaccounted for, source of 
uncertainty is that due to the uncertainty in the modelling of various 
physical effects.

Table 8 list the $P(k)$ normalization amplitudes $A$ (e.g., eqs. [1] and [2]) 
when $Q_{\rm rms-PS} = 10\ \mu$K. Examples of the power spectra normalized to 
$Q_{\rm rms-PS}$ derived from the mean of the DMR four-year data analysis 
extreme upper and lower 2-$\sigma$ limits discussed above are shown in Figs. 
23. One will notice, from Fig. 23(e), the good agreement between the 
open-bubble inflation spectra. 

When normalized to the two extreme 2-$\sigma$ $Q_{\rm rms-PS}$ limits (e.g., 
cols. [5] and [8] of Table 10), the $P(k)$ normalization factor (eq. [1] and 
Table 8) for the open-bubble inflation model (RP94; BGT; YST), may be 
summarized by, for the lower 2-$\sigma$ limit,
$$
 {h^4 A(\Omega_0) \over 10^5\, {\rm Mpc}^4} \simeq 4.3 + 1.95\, 
  {\rm sin}\left[ 1.07\pi (\Omega_0 - 0.1)^{0.85} \right] ,
 \eqno(8)
$$
and for the upper 2-$\sigma$ limit,
$$
 {h^4 A(\Omega_0) \over 10^5\, {\rm Mpc}^4} \simeq 9.3 + 3.35\,
  {\rm sin}\left[ 1.13\pi (\Omega_0 - 0.1)^{0.78} \right] .
 \eqno(9)
$$
These fits are good to $\sim 1\%$ for $0.1 \leq \Omega_0 \leq 1$. Note however
that they are derived using the $Q_{\rm rms-PS}$ values determined for given
$t_0$ and $\Omega_B h^2$ and hence do not account for the additional 
uncertainty introduced by allowed variations in these parameters (which 
could affect the power spectrum normalization amplitude by as much as 
$\sim 3-4\%$). From Fig. 23(e), and given the uncertainties, we see that the
fitting formulae of eqs. (8) and (9) provide an adequate summary for all
the open-bubble inflation model spectra. 

The extreme $\pm 2$-$\sigma$ $P(k)$ normalization
factor (eq. [2] and Table 8) for the flat-space scale-invariant spectrum
open model (W83) may be summarized by, for the lower 2-$\sigma$ limit,
$$
 {h^4 A(\Omega_0) \over 10^5\, {\rm Mpc}^4} \simeq 4.85 + 2.9\,
  {\rm cos}\left[ 0.9\pi \vert\Omega_0 - 0.325\vert^{1.25} \right] ,
 \eqno(10)
$$
and for the upper 2-$\sigma$ limit,
$$
 {h^4 A(\Omega_0) \over 10^5\, {\rm Mpc}^4} \simeq 11 + 5\,
  {\rm cos}\left[ 0.85\pi \vert\Omega_0 - 0.2\vert^{1.2} \right] .
 \eqno(11)
$$
These fits are good to better than $\sim 2\%$ for $0.2 \leq \Omega_0 \leq 1$;
again, they are derived from $Q_{\rm rms-PS}$ values determined at given
$t_0$ and $\Omega_B h^2$.

Given the uncertainties involved in the normalization procedure (born of both 
statistical and other arguments) it is not yet possible to quote a unique DMR
normalization amplitude (G96). As a  ``central" value for the $P(k)$
normalization factor, we currently advocate the mean of eqs. (8) and (9) or 
eqs. (10) and (11) as required. We emphasize, however, that it is incorrect to 
draw conclusions about model viability based solely on this ``central" value. 

In conjunction with numerically determined transfer functions, the fits of 
eqs. (8)--(11) allow for a determination of \sgm, accurate to a few percent. 
Here the mean square linear mass fluctuation averaged over a sphere of 
coordinate radius $\bar\chi$ is
\setcounter{equation}{11}
\begin{eqnarray}
   \left\langle\left[{\delta M\over M} (\bar\chi) \right]^2\right\rangle 
 & = &
   {2 \over \pi^2 \left[{\rm sinh} (\bar\chi)\, {\rm cosh}(\bar\chi)
    - \bar\chi\right]^2}
   \nonumber \\
   {\ }
 & {\ } &
   \times \int^\infty_0 {dk \over (1 + k^2)^2}
   \left[{\rm cosh}(\bar\chi)\, {\rm sin}(k\bar\chi) - k\, {\rm sinh}(\bar\chi)
   \, {\rm cos}(k\bar\chi) \right]^2 P(k) ,
\end{eqnarray}
which, on small scales, reduces to the usual flat-space expression 
$[9/2\pi^2] \int^\infty_0
dk\, k^2 P(k) \left[{\rm sin}(k\bar\chi) - k\bar\chi \, {\rm cos}(k\bar\chi)
\right]^2/(k\bar\chi)^6$. 

If instead use is made of the Bardeen et al. (1986, hereafter BBKS) analytic 
fit to the transfer function using the parameterization of eq. (13) below 
(Sugiyama 1995) and numerically determined values for $A$, the resultant 
$(\delta M/M)[8h^{-1}\ {\rm Mpc}]$ values are accurate to better than 
$\sim 5\%$ (except for large baryon-fraction, $\Omega_B/\Omega_0 \gap 0.4$, 
models where the error could be as large as $\sim 7\%$). Use of the analytic 
fits of eqs. (8)--(11) for $A$ (instead of the numerically determined values) 
slightly increases the error, while use of the BBKS transfer function fit 
parameterized by an earlier version of eq. (13) below, $S = \Omega_0 h
\, {\rm exp}\left[ -\Omega_B (1+\Omega_0)/\Omega_0\right]$, results in 
$(\delta M/M)[8 h^{-1}\ {\rm Mpc}]$ values that could be off by as much
as $\sim 7-10\%$. Nevertheless, as has been demonstrated by LLRV, the 
approximate analytic fit to the transfer function greatly simplifies the 
computation and allows for rapid demarcation of the favoured part of 
cosmological-parameter space.

Numerical values for some cosmographic and large-scale structure statistics
for the models considered here are recorded in Tables 9--15. We emphasize 
that when comparing to observational data we make use of 
numerically-determined large-scale structure predictions, not those derived 
using an approximate analytic fitting formula. 

Tables 9--12 give the predictions for the open-bubble inflation model accounting
only for the perturbations generated during the evolution inside the bubble
(RP94), and for the flat-space scale-invariant spectrum 
open model (W83). Each of these tables corresponds to a different 
pair of $(t_0,\ \Omega_B h^2)$ values. The first two columns in these tables 
record $\Omega_0$ and $h$, and the third column is the cosmological
baryonic-matter fraction $\Omega_B/\Omega_0$. The fourth column gives the 
value of the matter power spectrum scaling parameter (Sugiyama 1995),
$$
   S = \Omega_0 h e^{-\Omega_B (\sqrt{2h} + \Omega_0)/\Omega_0} ,
   \eqno(13)
$$
which is used to parameterize approximate analytic fits to the power spectra 
derived from numerical integration of the perturbation equations. The 
quantities listed in columns (1)--(4) of these tables are sensitive only to 
the global parameters of the cosmological model.

Columns (5) and (8) of Tables 9--12 give the DMR data 2-$\sigma$ range of 
$Q_{\rm rms-PS}$ that is used to normalize the perturbations in the models
considered here. The numerical values in Table 12 are for 
$t_0 \simeq 12$ Gyr, $\Omega_B h^2 = 0.007$. We did not analyze the DMR data
using $C_\ell$'s for these models, and in this case the perturbations are 
normalized to the $Q_{\rm rms-PS}$ values from the $t_0 \simeq 10.5$ Gyr,
$\Omega_B h^2 = 0.0055$ analyses. (As discussed above, shifts in $h$ and 
$\Omega_B h^2$ do not greatly alter the inferred normalization amplitude.)

Columns (6) and (9) of Tables 9--12 give the 2-$\sigma$ range of 
$(\delta M/M)[8h^{-1} \ {\rm Mpc}]$. These were determined using the 
$P(k)$ derived from numerical integration of the perturbation equations. 
For about two dozen cases, these rms mass fluctuations determined using
the two independent numerical integration codes were compared and found
to be in excellent agreement. (At fixed $Q_{\rm rms-PS}$, they differ by 
$\sim 0.002-0.5\%$ depending on model-parameter values, with the typical 
difference being $\sim 0.1\%$. We again emphasize that this is mostly a 
reflection of currently achievable numerical accuracy.). 

To usually better than $\sim 3\%$ accuracy, for $0.2 \leq \Omega_0
\leq 1$, the 2-$\sigma$ $(\delta M/M) [8h^{-1}\ {\rm Mpc}]$ entries 
of columns (6) and (9) of Tables 9--12 may be summarized by the fitting
formulae listed in Table 14. These fitting formulae are
more accurate than expressions for $(\delta M/M)[8 h^{-1}\ {\rm Mpc}]$
derived at the same cosmological-parameter values using an analytic 
approximation to the transfer function and the normalization of eqs. (8)--(11).

For open models, as discussed below, it proves most convenient to 
characterize the peculiar velocity perturbation by the parameter
$$
   \beta_I = {\Omega_0{}^{0.6} \over b_{IRAS}} =
       1.3 \Omega_0{}^{0.6} {\delta M \over M}(8h^{-1}\ {\rm Mpc}) ,
   \eqno(14)
$$
where $b_{IRAS}$ is the linear bias factor for $IRAS$ galaxies (e.g., Peacock
\&\ Dodds 1994). The 2-$\sigma$ range of $\beta_I$ are listed in columns (7) 
and (10) of Tables 9--12. 

Table 13 compares the $(\delta M/M)[8h^{-1}\ {\rm Mpc}]$ values for spectra 
of types (1)--(3) above. Clearly, there is no significant observational 
difference between the predictions for the different spectra. In what follows, 
for the open-bubble inflation model we concentrate on the type (1) spectrum 
above. 

Again, the ranges in Tables 9--14 are those determined from the maximal
2-$\sigma$ $Q_{\rm rms-PS}$ range. Table 15 lists ``central DMR-normalized"  
values for $(\delta M/M) [8h^{-1}\ {\rm Mpc}]$, defined as the mean of
the maximal $\pm$2-$\sigma$ entries of Tables 9--12. (The mean of the 
$\pm$2-$\sigma$ fitting formulae of Table 14 may be used to interpolate
between the entries of Table 15.) We again emphasize that it is incorrect
to draw conclusions about model viability based solely on these ``central"
values --- for the purpose of constraining model-parameter values by, e.g., 
comparing numerical simulation results to observational data one must 
make use of computations at a few different values of the normalization
selected to span the $\pm$2-$\sigma$ ranges of Tables 9--12. 

\section{CURRENT OBSERVATIONAL CONSTRAINTS ON DMR-NORMALIZED MODELS}

The DMR likelihoods do not meaningfully exclude any part of the ($\Omega_0$, 
$h$, $\Omega_B h^2$) parameter space for the models considered here. In this 
section we combine current observational constraints on global cosmological 
parameters with the DMR-normalized model predictions to place constraints on 
the range of allowed model-parameter values. It is important to bear in mind 
that some measures of observational cosmology remain uncertain thus our 
analysis here must be viewed as tentative and subject to revision as the 
observational situation approaches equilibrium. To constrain our 
model-parameter values we have employed the most robust of the current 
observational constraints. Tables 9--12 list some observational predictions 
for the models considered here, and the boldface entries are those that are 
inconsistent with current observational data at the 2-$\sigma$ significance 
level. 

\subsection{Observational Constraints Used}

For each cosmographic or large-scale parameter, we have generally chosen to
use constraints from a single set of observations or from a single analysis.
We generally use the most recent analyses since we assume that they 
incorporate a better understanding of the uncertainties, especially those due 
to systematics. The specific constraints we use are summarized below, where 
we compare them to those derived from other analyses.

The model predictions depend on the age of the universe $t_0$. To reconcile 
the models with the high measured values of the Hubble parameter
$h$, we have chosen to focus on $t_0 \simeq$ 10.5, 12, and 13.5 Gyr, which
are near the lower end of the ages now under discussion. For instance, Jimenez 
et al. (1996) find that the oldest globular clusters have ages $\sim 11.5-15.5$
Gyr (also see Salaris, Degl'Innocenti, \&\ Weiss 1996; Renzini et al. 1996), 
and that it is very unlikely that the oldest clusters are younger than 9.7 Gyr.

The value of $\Omega_0$ is another input parameter for
our computations. As summarized by Peebles (1993, $\S 20$), on scales $\lap
10h^{-1}$ Mpc a variety of different observational measurements indicate that
$\Omega_0$ is low. For instance, virial analyses of X-ray cluster data 
indicates $\Omega_0 = 0.24$, with a 2-$\sigma$ range: $0.04 < \Omega_0 
< 0.44$ (Carlberg et al. 1996 --- we have added their 1-$\sigma$ statistical
and systematic uncertainties in quadrature and doubled to get the 2-$\sigma$ 
uncertainty). In a CDM model in which structure forms at a relatively high
redshift (as is observed), these local estimates of $\Omega_0$ do constrain
the global value of $\Omega_0$ (since, in this case, it is inconceivable that 
the pressureless CDM is much more homogeneously distributed than is the 
observed baryonic mass). We hence adopt a 2-$\sigma$ upper limit of $\Omega_0
< 0.6$ to constrain the CDM models we consider here. (This large upper limit 
allows for the possibility that the models might be moderately biased.)
The boldface entries in column (1) of Tables 9--12 indicates those 
$\Omega_0$ values inconsistent with this constraint.

Column (2) of Tables 9--12 gives the value of the Hubble parameter $h$ that
corresponds to the chosen values of $\Omega_0$ and $t_0$. Current 
observational data favours a larger $h$ (e.g., Kennicutt, Freedman, \&\
Mould 1995; Baum et al. 1995; van den Bergh 1995; Sandage et al. 1996; 
Ruiz-Lapuente 1996; Riess, Press, \&\ Kirshner 1996; but also see Schaefer 
1996; Branch et al. 1996). For the purpose of our analysis here we adopt the 
$HST$ value $h = 0.69 \pm 0.08$ (1-$\sigma$ uncertainty, Tanvir et al. 1995); 
doubling the uncertainty, the 2-$\sigma$ range is $0.53 \leq h \leq 0.85$. 
The bold face entries in column (2) of Tables 9--12 indicates those 
model-parameter values which predict an $h$ inconsistent with this range.

Comparison of the standard nucleosynthesis theoretical predictions for the 
primordial light element abundances to what is determined by extrapolation of
the observed abundances to primordial values leads to constraints on 
$\Omega_B h^2$. It has usually been argued that ${}^4$He and ${}^7$Li allow
for the most straightforward extrapolation from the locally observed abundances
to the primordial values (e.g., Dar 1995; Fields \&\ Olive 1996; Fields et al.
1996, hereafter FKOT). The observed ${}^4$He and ${}^7$Li abundances then 
suggest $\Omega_B h^2 = 0.0066$, and a conservative assessment of the 
uncertainties indicate a 2-$\sigma$ range: $0.0051 < \Omega_B h^2 < 0.016$ 
(FKOT; also see Copi et al. 1995; Sarkar 1996).

Observational constraints on the primordial deuterium (D) abundance should, 
in principle, allow for a tightening of the allowed $\Omega_B h^2$ range. 
There are now a number of different estimates of the primordial D abundance, 
and since the field is still in its infancy it is, perhaps, not surprising 
that the different estimates are somewhat discrepant. Songaila et al. (1994), 
Carswell et al. (1994), and Rugers \&\ Hogan (1996a,b) use observations of 
three high-redshift absorption clouds to argue for a high primordial
D abundance and so a low $\Omega_B h^2$. Tytler, Fan, \&\ Burles (1996) and
Burles \&\ Tytler (1996) study two absorption clouds and argue for a low 
primordial D abundance and so a high $\Omega_B h^2$. Carswell et al. (1996)
and Wampler et al. (1996) examine other absorption clouds, but are not able to 
strongly constrain $\Omega_B h^2$.
While the error bars on $\Omega_B h^2$ determined from these D abundance
observations are somewhat asymmetric, to use these results to qualitatively 
pick the $\Omega_B h^2$ values we wish to examine we assume that the errors are 
gaussian (and where needed add all uncertainties in quadrature
to get the 2-$\sigma$ uncertainties). The large D abundance 
observations suggest $\Omega_B h^2 = 0.0062$
with a 2-$\sigma$ range: $0.0046 < \Omega_B h^2 < 0.0078$ (Rugers \&\ Hogan 
1996a). When these large D abundances are combined with the observed ${}^4$He
and ${}^7$Li abundances, they indicate $\Omega_B h^2 = 0.0064$, with a 
2-$\sigma$ range: $0.0055 < \Omega_B h^2 < 0.0087$ (FKOT). 
The large D abundances are consistent with the standard interpretation of the 
${}^4$He and ${}^7$Li abundances, and with the standard model of particle 
physics (with three massless neutrino species); they do, however, seem to 
require a modification in galactic chemical evolution models to be consistent 
with local determinations of the D and ${}^3$He abundances (e.g., FKOT; 
Cardall \&\ Fuller 1996). The low D abundance observations favour 
$\Omega_B h^2 = 0.024$ with a 2-$\sigma$ range: $0.018 < \Omega_B h^2
< 0.030$ (Burles \&\ Tytler 1996). The low D abundance observations seem to 
be more easily accommodated in modifications of the standard model of particle 
physics, i.e., they are difficult to reconcile with exactly three massless 
neutrino species; alternatively they might indicate a gross, as yet unaccounted 
for, uncertainty in the observed ${}^4$He abundance (Burles \&\ Tytler 1996; 
Cardall \&\ Fuller 1996). The low D abundance is approximately consistent
with locally-observed D abundances, but probably requires some modification 
in the usual galactic chemical evolution model for ${}^7$Li (Burles \&\ Tytler
1996; Cardall \&\ Fuller 1996).

To accommodate the range of $\Omega_B h^2$ now under discussion, we compute 
model predictions for $\Omega_B h^2 = 0.0055$ (Table 9), 0.007 (Table 12), 
0.0125 (Table 10), and 0.0205 (Table 11). We shall find that this uncertainty 
in $\Omega_B h^2$ precludes determination of robust constraints on 
model-parameter values. Fortunately, recent improvements in observational 
capabilities
should eventually lead to a tightening of the constraints on $\Omega_B h^2$,
and so allow for tighter constraints on the other cosmological parameters.

Column (3) of Tables 9--12 give the cosmological baryonic-mass fraction 
for the models we consider here. The cluster baryonic-mass fraction is the
sum of the cluster galactic-mass and gas-mass fractions. Assuming that the 
White et al. (1993) 1-$\sigma$ uncertainties on the cluster total, galactic,
and gas masses are gaussian and adding them in quadrature, we find
for the 2-$\sigma$ range of the cluster baryonic-mass fraction:
$$
   {M_B \over M_{\rm total}} = (1 \pm 0.55) \left( 0.009 + {0.05 \over h^{1.5}}
      \right) .
   \eqno(15)
$$
Elbaz, Arnaud, \&\ B\"ohringer (1995), White \&\ Fabian (1995), David, Jones,
\&\ Forman (1995), Markevitch et al. (1996), and Buote \&\ Canizares (1996)
find similar (or larger) gas-mass fractions. Note that Elbaz et al. (1995)
and White \&\ Fabian (1995) find that the gas-mass error bars are somewhat 
asymmetric; this non-gaussianity is ignored here. Assuming that the cluster
baryonic-mass fraction is an unbiased estimate of the cosmological baryonic-mass
fraction, we may use eq. (15) to constrain the cosmological parameters. The 
boldface entries in column (3) of Tables 9-12 indicates those model-parameter
values which predict a cosmological baryonic-mass fraction inconsistent
with the range of eq. (15).

Viana \&\ Liddle (1996, hereafter VL) have reanalyzed the combined galaxy
$P(k)$ data of Peacock \&\ Dodds (1994), ignoring some of the smaller scale 
data where nonlinear effects might be somewhat larger than previously 
suspected. Using an analytic approximation to the $P(k)$, they estimate
that the scaling parameter (eq. [13])\footnote{
VL actually set $2 h = 1$ in the exponent of eq. (13), so the numerical values
of their constraint on $S$ should be reduced slightly. We ignore this small
effect here.} 
$S = 0.23$, with a 2-$\sigma$ range,
$$
    0.20 \leq S \leq 0.27 .
    \eqno(16)
$$
This estimate is consistent with earlier ones\footnote{
LLRV used results from an earlier analysis which favoured larger 
values of $S$ than eq. (16) --- this is one reason why LLRV favour a higher 
$\Omega_0$ for the open-bubble inflation model than do GRSB.}.
It might be of interest
to determine whether the wiggles in $P(k)$ due to the pressure in the 
photon-baryon fluid, see Figs. 23, can significantly affect the determination 
of $S$, especially in large $\Omega_B/\Omega_0$ models. (These wiggles are not
well described by the analytic approximation to $P(k)$.) The boldface entries 
in column (4) of Tables 9--12 indicates those model-parameter values which 
predict a scaling parameter value inconsistent with the range of eq. (16).

To determine the value of the linear bias parameter $b$,
$$
   {\delta N \over N}(8 h^{-1}\ {\rm Mpc}) = b {\delta M \over M} 
      (8 h^{-1}\ {\rm Mpc}) ,
   \eqno(17)
$$
where $\delta N/N$ is the rms fractional perturbation in galaxy number, we
adopt the APM value (Maddox, Efstathiou, \&\ Sutherland 1996) of 
$(\delta N/N)[8 h^{-1}\ {\rm Mpc}] = 0.96$, with 2-$\sigma$ range:
$$
   0.75 < {\delta N \over N}(8 h^{-1}\ {\rm Mpc}) < 1.2 ,
   \eqno(18)
$$
where we have added the uncertainty due to the assumed cosmological model and 
due to the assumed evolution in quadrature with the statistical 1-$\sigma$ 
uncertainty (Maddox et al. 1996, eq. [43]), and doubled to get the
2-$\sigma$ uncertainty. The range of eq. (18) is consistent with that 
determined from eqs. (7.33) and (7.73) of Peebles (1993).

The local abundance of rich clusters, as a function of their X-ray 
temperature, provides a tight constraint on $(\delta M/M)[8h^{-1}\ {\rm Mpc}]$.
Eke, Cole, \&\ Frenk (1996, hereafter ECF) (and 
S. Cole, private communication 1996) find for the open model at 2-$\sigma$:
$$
   {\delta M \over M} (8h^{-1}\ {\rm Mpc}) = (0.52 \pm 0.08) 
     \Omega_0{}^{-0.46 + 0.10\Omega_0} ,
   \eqno(19)
$$
where we have assumed that the ECF uncertainties are gaussian\footnote{
Note that the constraint of eq. (19) is that derived for a fixed $S$, and that 
in general it depends weakly on the value of $S$ (and so on the value of $h$
and $\Omega_B$) --- see Fig. 13 of ECF. In our preliminary 
analysis here we ignore this mild dependence on $h$ and $\Omega_B$. Also note
that the constraint of eq. (19) is approximately that required for 
consistency with the observed cluster correlation function.}.
The constraints of eq. (19) are consistent with, but more restrictive than,
those derived by VL\footnote{
VL favour $(\delta M/M)[8 h^{-1}\ {\rm Mpc}]
= 0.60$ for fiducial CDM, which is at the $+$2-$\sigma$ limit of eq. (19).
(As discussed in ECF, this is because VL normalize to the cluster
temperature function at 7 keV, where there is a rise in the temperature 
function.) This is one reason why LLRV favour a higher value 
of $\Omega_0$ for the open-bubble inflation model than did GRSB.}.
This is because ECF use observational data over a larger range in X-ray 
temperature to constrain $\delta M/M$, and also use N-body computations at 
$\Omega_0 =$ 0.3 and 1 to calibrate the Press-Schechter model (which is used 
in their determination of the constraints). Furthermore, ECF also make use of 
hydrodynamical simulations of a handful of individual clusters in the fiducial
CDM model ($\Omega_0 = 1$) to calibrate the relation between the gas 
temperature and the cluster mass, and then use this calibrated relation for 
the computations at all values of $\Omega_0$. The initial conditions for all the
simulations were set using the analytical approximation to $P(k)$, so again 
it might be of interest to see whether the wiggles in the numerically 
integrated $P(k)$ could significantly affect the determination of the 
constraints of eq. (19). Kitayama \&\ Suto (1996) use X-ray cluster data, and 
a method that allows for the fact that clusters need not have formed at the 
redshift at which they are observed, to directly constrain the value of 
$\Omega_0$ for CDM cosmogonies normalized by the DMR two-year data. Their 
conclusions are in resonable accord with what would be found by using eq. (19) 
(derived assuming that observed clusters are at their redshifts of formation). 
However, Kitayama \&\ Suto (1996) note that 
evolution from the redshift of formation to the redshift of observation can 
affect the conclusions, so a more careful comparison of these two results is
warranted. The boldface entries in columns (6) and (9) of Tables 9--12 
indicate those model-parameter values whose predictions are inconsistent with 
the constraints of eq. (19)\footnote{
Given the $\sim \pm 8\%$ (1-$\sigma$) uncertainty of eq. (19), approximate
analyses based on using the analytic BBKS approximation to the transfer 
function should make use of the more accurate parameterization of eq. (13) 
(rather than that with $2h = 1$ in the exponent), as this gives 
$(\delta M/M)[8 h^{-1}\ {\rm Mpc}]$ to better than $\sim 5\%$ in the 
observationally viable part of parameter space (provided use is made of
the numerically determined values of $A$).}. 

From large-scale peculiar velocity observational data Zaroubi et al. (1996) 
estimate $(\delta M/M)[8 h^{-1}\ {\rm Mpc}] = (0.85 \pm 0.2)\Omega_0{}^{-0.6}$ 
(2-$\sigma$). It might be significant that the large-scale peculiar velocity 
observational data constraint is somewhat discordant with (higher than) the 
cluster temperature function constraint.

Since $J_3$ is less sensitive to smaller length scales (compared to 
$(\delta M/M)[8 h^{-1}\ {\rm Mpc}]$), observational constraints on $J_3$ are
more reliably contrasted with the linear theory predictions. However, since
$J_3$ is sensitive to larger length scales, the observational constraints on
$J_3$ are significantly less restrictive than the $\pm 8\%$ (1-$\sigma$)
constraints of eq. (19), and so we do not record the predicted values of $J_3$
here. 

Observational constraints on the mass power spectrum determined from 
large-scale peculiar velocity observations provide another constraint on the 
mass fluctuations. Kolatt \&\ Dekel (1995) find at the 1-$\sigma$ level
$$
   h^3 P(k/h = 0.1 \ {\rm Mpc}^{-1}) = (4.6 \pm 2.3) \times 10^3 
       \Omega_0{}^{-1.2}\ {\rm Mpc}^3 ,
   \eqno(20)
$$
where the 1-$\sigma$ uncertainty also accounts for sample variance
(T. Kolatt, private communication 1996). Since the uncertainties associated 
with the constraint of eq. (19) are more restrictive than those 
associated with the constraint of eq. (20), we do not tabulate predictions for
this quantity here. However, comparison may be made to the 
predicted linear theory mass power spectra of Figs. 23, bearing in mind the 
$\sim \pm 4.6$ (2-$\sigma$) uncertainty of eq. (20) (the uncertainty is 
approximately gaussian, T. Kolatt, private communication 1996),\footnote{
Thus at the higher, $\sim 2$-$\sigma$, significance level, eq. (20) 
provides a strong upper limit on $P(k/h = 0.1\ {\rm Mpc}^{-1})$, especially
at larger $\Omega_0$ because of the $\Omega_0$ dependence.}
and the uncertainty in the DMR normalization (not shown in Figs. 23).

Columns (7) and (10) of Tables 9--12 give the DMR-normalized model predictions
for $\beta_I$ (eq. [14]). Cole, Fisher, \&\ Weinberg (1995) measure the 
anisotropy of the redshift space power spectrum of the $IRAS$ 1.2 Jy survey and 
conclude $\beta_I = 0.52$ with a 2-$\sigma$ c.l. range: 
$$
    0.24 \leq \beta_I \leq 0.80 ,
    \eqno(21)
$$
where we have doubled the error bars of eq. (5.1) of Cole et al. (1995) to 
get the 2-$\sigma$ range. Cole et al. (1995, Table 1) compare the estimate of
eq. (21) to other estimates of $\beta_I$, and at 2-$\sigma$ all estimates of
$\beta_I$ are consistent. It should be noted that the model predictions of 
$\beta_I$ (eq. [14]) in Tables 9--12 assume that for $IRAS$ galaxies 
$(\delta N/N)[8 h^{-1}\ {\rm Mpc}] = 1/1.3$ holds exactly, i.e., they ignore
the uncertainty in the rms fractional perturbation in $IRAS$ galaxy number,
which is presumably of the order of that in eq. (18). As the constraints from
the deduced $\beta_I$ values, eq. (21), are not yet as restrictive as those
from other large-scale structure measures, we do not pursue this issue in
our analysis here. The boldface entries in columns (7) and (10) of
Tables 9--12 indicate those model-parameter values whose predictions are 
inconsistent with the constraints of eq. (21).

\subsection{Constraints on Model-Parameter Values}      

The boldface entries in Tables 9--12 summarize the current constraints 
imposed by the observational data discussed in the previous section on
the model-parameter values for the open-bubble inflation model 
(spectra of type [1] above), and for the flat-space scale-invariant spectrum
open model (type [4] above). The current observational constraints on the 
models are not dissimilar, but this is mostly a reflection of the uncertainty 
on the constraints themselves since the model predictions are fairly
different.

In the following discussion of the preferred part of model-parameter space
we focus on the open-bubble inflation model (RP94). Note from Table 13 that 
the large-scale structure predictions of the open-bubble inflation model do 
not depend on perturbations generated in the first epoch of 
inflation (BGT; YST), and also do not depend significantly on the contribution 
from the non-square-integrable basis function (YST).

Table 9 corresponds to the part of parameter space with ``maximized" 
small-scale power in matter fluctuations. This is accomplished by 
picking a low $t_0 \simeq 10.5$ Gyr (and so large $h$), and by picking
a low $\Omega_B h^2 = 0.0055$ (this is the lower 2-$\sigma$ limit from 
standard nucleosynthesis and the observed ${}^4$He, ${}^7$Li, and high
D abundances, FKOT). 
The tightest constraints on the model-parameter values come from the 
matter power spectrum observational data constraints on the shape
parameter $S$ (Table 9, col. [4]), and from the cluster X-ray temperature 
function observational data constraints on $(\delta M/M)[8 h^{-1}\ {\rm Mpc}]$
(col. [6]). Note that for $\Omega_0 = 0.3$ the predicted upper 2-$\sigma$ 
value of $(\delta M/M)[8h^{-1}\ {\rm Mpc}] = 0.69$, while ECF
conclude that at 2-$\sigma$ the observational data requires that this be
at least 0.74, so an $\Omega_0 = 0.3$ case fails this test. The 
constraints on $\beta_I$ (col. [7]) are not as restrictive as those on 
$(\delta M/M)[8h^{-1}\ {\rm Mpc}]$. For these values of $t_0$ and $\Omega_B
h^2$ the cosmological baryonic-mass fraction at $\Omega_0 = 0.3$ is predicted 
to be 0.033 (col. [3]), while at 2-$\sigma$ White et al. (1993) require that 
this be at least 0.039 (at $h = 0.75$), so again this $\Omega_0 = 0.3$ model 
just fails this test. Given the observational uncertainties, it might be 
possible to make minor adjustments to model-parameter values so that an 
$\Omega_0 \sim 0.3-0.35$ model with $t_0 \sim 10.5$ Gyr and $\Omega_B h^2 
\sim 0.0055$ is just consistent with the observational data. However, it is 
clear that current observational data do not favour an open model with 
$\Omega_B h^2 \simeq 0.0055$ --- the observed cluster $(\delta M/M)[8h^{-1}\ 
{\rm Mpc}]$ favours a larger $\Omega_0$ while the observed cluster 
baryonic-mass fraction favours a smaller $\Omega_0$, and so are in conflict.

Table 10 gives the predictions for the $t_0 \simeq 12$ Gyr, $\Omega_B h^2
= 0.0125$ models. This value of $\Omega_B h^2$ is consistent with
the 2-$\sigma$ range determined from standard nucleosynthesis and the observed
${}^4$He and ${}^7$Li abundances: $0.0051 < \Omega_B h^2 < 0.016$ (FKOT,
also see Copi et al. 1995; Sarkar 1996). It is, however, somewhat 
difficult to reconcile $\Omega_B h^2 = 0.0125$ with the 2-$\sigma$ range 
derived from the observed ${}^4$He, 
${}^7$Li, and current high D abundances $0.0055 < \Omega_B h^2 < 0.0087$
(FKOT), or with that from the current observed low D abundances
$0.018 < \Omega_B h^2 < 0.030$ (Burles \&\ Tytler 1996). In any case, the 
observed D abundances are still under discussion, and must be viewed as 
preliminary. In this case, open-bubble inflation models with $ 0.35 < \Omega_0
\lap 0.5$ are consistent with the observational constraints. The current 
central observational data values for $S$ and $\beta_I$ favour $\Omega_0 \sim
0.4$, while that for the cluster baryonic-mass fraction prefers $\Omega_0 
\sim 0.3$, and that for $(\delta M/M) [8h^{-1}\ {\rm Mpc}]$ favours
$\Omega_0 \sim 0.45$, so in this case the agreement between predictions and 
observational data is fairly impressive (although the Tanvir et al. 1995 central
$h$ value favours $\Omega_0 \sim 0.2$). Note that in this case models with 
$\Omega_0 \gap 0.6$ are quite inconsistent with the data.

Table 11 gives the predictions for $t_0 \simeq 13.5$ Gyr, $\Omega_B h^2 = 
0.205$ models. This baryonic-mass density value is consistent with that 
determined from the current observed low D abundances, but is difficult to 
reconcile with the current standard nucleosynthesis interpretation of the 
observed ${}^4$He and ${}^7$Li abundances (Cardall \&\ Fuller 1996). The 
larger value of $\Omega_B h^2$ (and smaller value of $h$) has now lowered 
small-scale power in mass fluctuations somewhat significantly, opening up the 
allowed $\Omega_0$ range to larger values. Models with $0.4 < \Omega_0
< 0.6$ are consistent with the observational data, although the higher 
$\Omega_0$ part of the range is starting to conflict with what is determined 
from the small-scale dynamical estimates, and the models do require a somewhat
low $h$ (but not yet inconsistently so at the 2-$\sigma$ significance level
--- while the Tanvir et al. 1995 central $h$ value requires $\Omega_0 < 0.1$,
at 2-$\sigma$ the $h$ constraint only requires $\Omega_0 \lap 0.6$). 
The central observational values for $S$, the cluster baryonic-mass fraction, 
$(\delta M/M)[8 h^{-1}\ {\rm Mpc}]$, and $\beta_I$ favour $\Omega_0 \sim 0.5$,
so the agreement with observational data is fairly impressive, and could even 
be improved by reducing $t_0$ a little to raise $h$.    

Table 12 gives the predictions for another part of model-parameter space. 
Here we show $\Omega_B h^2 = 0.007$ models (at $t_0 \simeq 12$ Gyr), consistent
with the central value of $\Omega_B h^2$ determined from standard 
nucleosynthesis using the observed ${}^4$He, ${}^7$Li, and high D abundances
(FKOT). The larger value of $\Omega_B h^2$ (compared to Table 9) eases the 
cluster
baryonic-mass fraction constraint, which now requires only $\Omega_0 < 0.4$.
The increase in $\Omega_B h^2$ also decreases the mass fluctuation amplitude,
making it more difficult to argue for $\Omega_0 = 0.3$; however, models
with $0.35 < \Omega_0 < 0.4$ seem to be consistent with the observational 
constraints when $\Omega_B h^2 \sim 0.007$ and $t_0 \sim 12$ Gyr. It is 
interesting that in this case the central observational data values we 
consider for $S$, for $(\delta M/M)[8 h^{-1}\ {\rm Mpc}]$, and for $\beta_I$ 
prefer $\Omega_0 \sim 0.4$; however,
that for the cluster baryonic-mass fraction (as well as that for $h$) favours 
$\Omega_0 \sim 0.2$ (although at 2-$\sigma$ the cluster baryonic-mass 
fraction constraint only requires $\Omega_0 < 0.4$). Hence, while
$\Omega_0 \sim 0.35-0.4$ open-bubble inflation models with $\Omega_B h^2
\sim 0.007$ and $t_0 \sim 12$ Gyr are quite consistent with the observational
constraints, in this case the agreement between predictions and observations 
is not spectacular. Note that in this case models with $\Omega_0 \gap 0.5-0.6$
are quite inconsistent with the observational data.
     
In summary, open-bubble inflation models based on the CDM picture (RP94; BGT;
YST) are reasonably 
consistent with current observational data provided $0.3 < \Omega_0 \lap 0.6$.
The flat-space scale-invariant spectrum open model (W83) is also 
reasonably compatible with current observational constraints for a similar 
range of $\Omega_0$. The uncertainty in current estimates of $\Omega_B h^2$
is one of the major reasons why such a large range in $\Omega_0$ is consistent
with current observational constraints. 

Our previous analysis of the DMR two-year data led us to conclude that only 
those open-bubble inflation models near the lower end of the above range 
($\Omega_0 \sim 0.3-0.4$) were consistent with the majority of observations
(GRSB). The increase in the allowed range to higher $\Omega_0$ values 
$\sim 0.5-0.6$ can be ascribed to a number of small effects. Specifically, 
these are: (1) the slight downward shift in the central value of the 
DMR four-year normalization relative to the two-year one (G96); (2) use of 
the full 2-$\sigma$ range of normalizations allowed by the DMR data analysis 
(instead of the 1-$\sigma$ range allowed by the galactic-frame 
quadrupole-excluded DMR two-year data set used previously);
(3) use of the 2-$\sigma$ range of the small-scale dynamical estimates
of $\Omega_0$ instead of the 1-$\sigma$ range used in our earlier analysis;
(4) we consider a range of $\Omega_B h^2$ values here (in GRSB we focussed 
on $\Omega_B h^2 = 0.0125$); and (5) we consider a range of $t_0$ values 
here (in GRSB we concentrated on $t_0 \simeq 12$ Gyr). 
We emphasize, however, that the part of parameter space with $\Omega_0 \sim 
0.5-0.6$ is only favoured if $\Omega_B h^2$ is large ($> 0.02$), $h$ is 
low $(< 0.55$), and the small-scale dynamical estimates of $\Omega_0$ turn
out to be biased somewhat low. 

\subsection{Indications from Additional Observational Constraints}  

The observational results we have used to constrain model-parameter values in
the previous sections are the most robust currently available.
In addition, there are several other observational results which we do
not consider to be as robust, and any conclusions drawn from these
should be treated with due caution. In this section we summarize several 
of the more tentative constraints from more recent observations.

In our analysis of the DMR two-year data normalized models, we compared model 
predictions for the rms value of the smoothed peculiar velocity field to 
results from the analysis of observational data (Bertschinger et al. 1990). 
We do not do so again here since, given the uncertainties, the conclusions 
drawn in GRSB are not significantly modified. In particular, comparison of the 
appropriate quantities implies that we can treat the old 1-$\sigma$ upper 
limits essentially as 2-$\sigma$ upper limits for the four-year analysis.

In GRSB we used $\beta_I$ determined by Nusser \&\ Davis (1994), $0.2 < 
\beta_I < 1.0$ (2-$\sigma$), to constrain the allowed range of models to $0.2 
< \Omega_0 \lap 0.6$. Here we use the Cole et al. (1995) estimate, $0.24 < 
\beta_I < 0.80$ (2-$\sigma$), which, for the models of Table 10, requires 
$\Omega_0 > 0.25$. This value is just slightly below the lower limit 
($\Omega_0 \gap 0.3$) derived from the Bertschinger et al. (1990) results in 
GRSB. We hence conclude that the large-scale flow results of 
Bertschinger et al. (1990) indicates a lower 2-$\sigma$ limit on $\Omega_0$ 
that is about $\Delta\Omega_0 \sim 0.05$ higher than that suggested by the 
redshift-space distortion analysis of Cole et al. (1995).\footnote{
Note that the lower limit from the Bertschinger et al. (1990) analysis
is not as restrictive as that set by the cluster X-ray temperature function
data constraints on $(\delta M/M) [8h^{-1}\ {\rm Mpc}]$.}
We however strongly emphasize that the central value of the large-scale flow
results of Bertschinger et al. (1990) does favour a significantly larger value 
of $\Omega_0$ than the rest of the data we have considered here. Furthermore, 
as discussed in detail in GRSB, there is some uncertainty in how to properly 
interpret large-scale velocity data in the open models, particularly given the 
large sample variance associated with the measurement of a single bulk 
velocity (Bond 1996, also see LLRV). A more careful analysis, as well as more 
observational data, is undoubtedly needed before it will be possible to 
robustly conclude that the large-scale velocity data does indeed force one to 
consider significantly larger values of $\Omega_0$ than is favoured by the rest 
of the observational constraints (and hence rules out the models considered 
here).

It might be significant that on comparing the mass power spectrum deduced 
from a refined set of peculiar velocity observations to the galaxy power 
spectrum determined from the APM survey, Kolatt \&\ Dekel (1995) estimate that
for the optically-selected APM galaxies $\beta = 0.80$ with a 2-$\sigma$ range,
$$
   0.60 < \beta < 1.0 .
   \eqno(22)
$$
(Note that it has been argued that systematic uncertainties preclude a 
believable determination of $\beta_I$ from a comparison of the observed 
large-scale peculiar velocity field to the $IRAS$ 1.2 Jy galaxy distribution, 
Davis, Nusser, \&\ Willick 1996.) This range is consistent 
with other estimates now under discussion. The Stromlo-APM 
comparison of Loveday et al. (1996) indicates $\beta \simeq 0.48$, with a 
2-$\sigma$ upper limit of 0.75, while Baugh (1996) concludes that $\beta < 1.0$ 
(2-$\sigma$), and Ratcliffe et al. (1996) argue for $\beta = 0.55 \pm 0.12$.
Using the APM range for $(\delta N/N)[8 h^{-1}\ {\rm Mpc}]$, eq. (18), the
Kolatt \&\ Dekel (1995) estimate of $\beta$, eq. (22), may be converted to
an estimate of $\delta M/M$, and at 2-$\sigma$,
$$
   {\delta M \over M}(8 h^{-1} \ {\rm Mpc}) = (0.45-1.2) \Omega_0{}^{-0.6} .
   \eqno(23)
$$
It is interesting that at $\Omega_0 = 1$ the lower part of this range is 
consistent with that determined from the cluster X-ray  temperature function
data, eq. (19), although at lower $\Omega_0$ eq. (23) indicates a larger value 
then does eq. (19) because of the steeper rise to low $\Omega_0$.

Zaroubi et al. (1996) have constrained model-parameter values by comparing 
large-scale flow observations to that predicted in the DMR two-year data
normalized open-bubble inflation model. They conclude that the open-bubble
inflation model provides a good description of the large-scale flow
observations if, at 2-$\sigma$,
$$
    0.31 < \Omega_0 h < 0.44 .
    \eqno(24)
$$
From Table 12 we see 
that an open-bubble inflation model with $\Omega_0 = 0.45$ and $h = 0.62$
provides a good fit to all the observational data considered in $\S 5.1$.
For $h = 0.62$ Zaroubi et al. (1996) conclude that at 2-$\sigma$ $\Omega_0
> 0.5$ (eq. [24]), just above our value of $\Omega_0 = 0.45$. Since the 
Zaroubi et al. (1996) analysis does not account for the uncertainty in the
DMR normalization (T. Kolatt, private communication 1996), it is still 
unclear if the constraints from the large-scale flow observations are 
in conflict with those determined from the other data considered
here (and so rule out the open-bubble inflation model). It might also be
significant that on somewhat smaller length scales there is support for a 
smaller value of $\Omega_0$ from large-scale velocity field
data (Shaya, Peebles, \&\ Tully 1995).

The cluster peculiar velocity function provides an alternate mechanism for
probing the peculiar velocity field (e.g., Croft \&\ Efstathiou 1994; 
Moscardini et al. 1995; Bahcall \&\ Oh 1996). Bahcall \&\ Oh (1996) conclude 
that current observational data is well-described by an $\Omega_0 = 0.3$
flat-$\Lambda$ model with $h = 0.67$ and $(\delta M/M) [8 h^{-1}\ {\rm Mpc}]
= 0.67$. This normalization is somewhat smaller than that indicated by the 
DMR data (e.g., Ratra \&\ Sugiyama 1995). While Bahcall \&\ Oh (1996) did not 
compare the cluster peculiar velocity function data to the 
predictions of the open-bubble inflation model, approximate estimates indicate
that this data is consistent with the open-bubble inflation 
model predictions for the range of $\Omega_0$ favoured by the other data we 
consider in $\S\S 5.1, 5.2$ --- see the $(\delta M/M)[8 h^{-1}\ {\rm Mpc}]$
values for the allowed models in Tables 9--12. Bahcall \&\ Oh (1996) also 
note that it is difficult, if not impossible, to reconcile the cluster 
peculiar velocity observations with what is predicted in high density models 
like fiducial CDM and MDM.

At fixed $(\delta M/M)[8h^{-1}\ {\rm Mpc}]$, low-density cosmogonies form
structure earlier than high density ones. Thus observations of structure
at high redshift may be used to constrain the matter density. As benchmarks,
we note that scaling from the results of the numerical simulations of 
Cen \&\ Ostriker (1993), in a open model with $(\delta M/M)[8 h^{-1}\ {\rm Mpc}
] = 0.8$ galaxy formation peaks at a redshift $z_g \simeq 2.3$ when $\Omega_0
= 0.45$ and at $z_g \simeq 2.5$ when $\Omega_0 = 0.4$. Thus the open-bubble
inflation model is not in conflict with observational indications that the 
giant elliptical luminosity function at $z \sim 1$ is similar to that at the 
present (e.g., Lilly et al. 1995; Glazebrook et al. 1995; Im et al. 1996),
nor is it in conflict with observational evidence for 
massive galactic disks at $z \sim 1$ (Vogt et al. 1996). These models can also 
accommodate observational evidence of massive star-forming galaxies at $z \sim 
1.5$ (Cowie, Hu, \&\ Songaila 1995), as well as the significant peak at 
$z \sim 2.2$ in the number of galaxies as a function of (photometric) redshift 
found in the Hubble Deep Field (Gwyn \&\ Hartwick 1996), and it is not 
inconceivable that objects like the $z  = 2.7$ ``protogalaxy" 
candidate\footnote{
Note that if the velocity dispersion of the nearby foreground cluster 
has actually been significantly underestimated, the striking properties of 
this object could mostly be a consequence of gravitational lensing and it
would seem to be more reasonably interpreted as a massive star-forming
galaxy (Williams \&\ Lewis 1996).}
(Yee et al. 1996; Ellingson et al. 1996) can be produced in these models. It 
is, however, at present unclear whether the open-bubble inflation model can 
accommodate a substantial population of massive star-forming galaxies at 
$z \sim 3-3.5$ (Steidel et al. 1996; Giavalisco, Steidel, \&\ Macchetto 1996), 
and if there are many more examples of massive damped Lyman$\alpha$ 
systems\footnote{
These have many of the properties expected of young galaxies (Wolfe 1993;
Djorgovski et al. 1996, and references therein).}
like the one at $z = 4.4$ 
(e.g., Lu et al. 1996; Wampler et al. 1996; Fontana et al. 1996),
then, depending on the masses, these might be a serious problem for the 
open-bubble inflation model. On the other hand, the recent discovery of galaxy
groups at $z \sim 2.4$ (e.g., Francis et al. 1996; Pascarelle et al. 1996) 
probably do not pose a serious threat for the open-bubble inflation model, while
massive clusters at $z \sim 0.5-1$ (e.g., Luppino \&\ Gioia 1995; Pell\'o et
al. 1996) can easily be accommodated in the 
model. It should be noted that in adiabatic $\Omega_0 = 1$ models normalized 
to fit the present small-scale observations, e.g., fiducial CDM (with a 
normalization inconsistent with that from the DMR), or MDM, or tilted CDM
(without a cosmological constant), it is quite difficult, if not impossible,
to accommodate the above observational indications of early structure 
formation (e.g., Ma \&\ Bertschinger 1994; Ostriker \&\ Cen 1996).

With the recent improvements in observational capabilities, neoclassical 
cosmological tests hold great promise for constraining the world model.
It might be significant that current constraints from these tests are 
consistent with that region of the open-bubble inflation model parameter
space that is favoured by the large-scale structure constraints. These tests
include the $HST$ elliptical galaxy number counts test (Driver et al. 1996), 
an early application of the apparent magnitude-redshift test using Type Ia
supernovae (Perlmutter et al. 1996), as well as analyses of the rate of 
gravitational lensing of quasars by foreground galaxies (e.g., Torres \&\ 
Waga 1996; Kochanek 1996). It should be noted that these tests are also 
consistent with $\Omega_0 = 1$ models, and plausibly with a time-variable 
cosmological ``constant" dominated spatially-flat model (e.g., Ratra \&\ 
Quillen 1992; Torres \&\ Waga 1996), but they do put pressure on the 
flat-$\Lambda$ CDM model.    

Smaller-scale CMB spatial anisotropy measurements will eventually significantly
constrain the allowed range of model-parameter values. Fig. 24 compares the
1-$\sigma$ range of CMB spatial anisotropy predictions for a few representative
open-bubble inflation (as well as flat-space scale-invariant spectrum open)
models to available CMB spatial anisotropy
observational data. From a preliminary comparison of the predictions of 
DMR two-year data normalized open-bubble inflation models to available 
CMB anisotropy observational data, Ratra et al. (1995) concluded that the 
range of parameter space for the open-bubble inflation model that was favoured
by the other observational data was also consistent with the small-scale CMB
anisotropy data. This result was quantified by GRS, who also
considered open-bubble inflation models normalized to the $\pm$1-$\sigma$
values of the DMR two-year data (and hence considered open-bubble
inflation models normalized at close to the DMR four-year data value, see 
Figs. 5 and 6 of GRS). GRS discovered that
(given the uncertainties associated with the smaller-scale measurements) the 
1-$\sigma$ uncertainty in the value of the DMR normalization precludes 
determination of robust constraints on model-parameter values, although the 
range of model-parameter space for the open-bubble inflation model favoured 
by the analysis here was found to be consistent with the smaller-scale CMB
anisotropy observations, and $\Omega_0 \sim 0.1$ open-bubble inflation models
were not favoured by the smaller-scale CMB anisotropy observational data
(GRS, Figs. 5 and 6).\footnote{
The recent analysis of Hancock et al. (1996b) is generally consistent with
these results. They conclude that $\Omega_0 \sim 0.7$ is favoured, but even
at 1-$\sigma$ $0.3 \leq \Omega_0 \leq 1.7$ is allowed --- this broad range
is consistent with the conclusion of GRS that it is not yet 
possible to meaningfully constrain cosmological-parameter values from the 
CMB anisotropy data alone. Note also that Hancock et al. (1996b) do not 
consider the effects of the systematic shifts between the various 
DMR data sets, and also exclude a number of data points, e.g., the four
MSAM points and the MAX3 MUP point (which is consistent with the recent 
MAX5 MUP result, Lim et al. 1996), which do not disfavour a lower value of
$\Omega_0$ for the open-bubble inflation model (Ratra et al. 1995; GRS).}
A detailed analysis of the UCSB South
Pole 1994 CMB anisotropy data (Gundersen et al. 1995) by Ganga et al. (1996a)
reaches a similar conclusion: at 1-$\sigma$ (assuming a gaussian marginal 
probability distribution) the data favours open-bubble inflation models with 
$\Omega_0 < 0.5$, while at 2-$\sigma$ the UCSB South Pole 1994 data is 
consistent with the predictions of the open-bubble, flat-$\Lambda$, and 
fiducial CDM inflation models.

\section{DISCUSSION AND CONCLUSION}

We have compared the DMR 53 and 90 GHz sky maps to a variety of open model 
CMB anisotropy angular spectra in order to infer the normalization of these 
open cosmogonical models. Our analysis explicitly quantifies the small shifts 
in the inferred normalization amplitudes due to: (1) the small differences 
between the galactic- and ecliptic-coordinate sky maps; (2) the inclusion or 
exclusion of the $\ell = 2$ moment in the analysis; and, (3) the faint 
high-latitude Galactic emission treatment. 
We have defined a maximal 2-$\sigma$ uncertainty range based on the
extremal solutions of the normalization fits, and a maximal 1-$\sigma$
uncertainty range may be defined in a similar manner. For this maximal
1-$\sigma$ $Q_{\rm rms-PS}$ range the fractional 1-$\sigma$ uncertainty, 
at fixed $\Omega_B$ and $h$ (but depending on the assumed CMB anisotropy
angular spectrum and model-parameter values), ranges between
$\sim 10\%$ and $\sim 12\%$\footnote{
Note that the quoted 1-$\sigma$ (statistical and systematic) uncertainty
of BW (footnote 4, also see Bunn, Liddle, \&\ White 1996), $7.6\%$, is 
smaller than the DMR four-year data 1-$\sigma$ uncertainty estimated in, 
e.g., G96, Wright et al. (1996), and here. This is because we explicitly 
estimate the effect of all known systematic uncertainties for each assumed 
CMB anisotropy angular spectrum, and account for them, in the most conservative
manner possible, as small shifts. (In particular: we do not just account 
for the small systematic difference between the galactic- and ecliptic-frame 
maps; we do not assume that any of the small systematic differences lead 
to model-independent systematic shifts in the inferred $Q_{\rm rms-PS}$
values; and we do not add the systematic shifts in quadrature with the 
statistical uncertainty.) Since our accounting of the uncertainties is the 
most conservative possible, our conclusions about model-viability are the
most robust possible.}.
(Compare this to the $\sim 8\%$, 1-$\sigma$, uncertainty of eq. [19].)
Since part of this uncertainty is due to the small systematic shifts, the 
maximal 2-$\sigma$ fractional uncertainty is smaller than twice the maximal 
1-$\sigma$ fractional uncertainty. For the largest possible 2-$\sigma$ 
$Q_{\rm rms-PS}$ range defined above, the fractional uncertainty varies between 
$\sim 16\%$ and $\sim 19\%$. Note that
this accounts for intrinsic noise, cosmic variance, and effects (1)--(3)
above. Other systematic effects, e.g., the calibration uncertainty (Kogut
et al.  1996b), or the beamwidth uncertainty (Wright et al. 1994), are
much smaller than the effects we have accounted for here. It has also been
shown that there is negligible non-CMB contribution to the DMR data sets
from known extragalactic astrophysical foregrounds (Banday et al. 1996b).

By analyzing the DMR maps using CMB anisotropy spectra at fixed $\Omega_0$ but
different $h$ and $\Omega_B$, we have also explicitly quantified the small 
shifts in the inferred normalization amplitude due to shifts in $h$ and
$\Omega_B$. Although these shifts do depend on the value of $\Omega_0$ 
and the assumed model power spectrum, given the other uncertainties, it is
reasonable to ignore these small shifts when normalizing the models
considered in this work.

We have analyzed the open-bubble inflation model, accounting only for the 
fluctuations generated during the evolution inside the bubble (RP94), 
including the effects of the fluctuations generated in the first epoch of 
spatially-flat inflation (BGT; YST), and finally accounting for the 
contribution from a non-square-integrable basis function (YST). For 
observationally viable open-bubble models, the observable predictions do not 
depend significantly on the latter two sources of anisotropy. The observable 
predictions of the open-bubble inflation scenario seem to be robust --- it 
seems that only those fluctuations generated during the evolution inside the 
bubble need to be accounted for.

As discussed in the Introduction, a variety of more specific realizations
of the open-bubble inflation scenario have recently come under scrutiny.
These are based on specific assumptions about the vacuum state prior
to open-bubble nucleation. In these specific realizations of the 
open-bubble inflation scenario there are a number of additional mechanisms for 
stress-energy perturbation generation (in addition to those in the models 
considered here), including those that come from fluctuations in the bubble 
wall, as well as effects associated with the nucleation of a nonzero size 
bubble. While current analyses suggest that such effects also do not add a 
significant amount to the fluctuations generated during the evolution inside 
the bubble, it is important to continue to pursue such investigations --- both 
to more carefully examine the robustness of the open-bubble inflation 
scenario predictions, as well as to try to find a reasonable particle
physics based realization of the open-bubble inflation scenario.

As has been previously noted for other CMB anisotropy angular spectra 
(G96), the various different DMR data sets lead to slightly different 
$Q_{\rm rms-PS}$ normalization amplitudes, but well within the statistical
uncertainty. This total range is slightly reduced if one considers results
from analyses either ignoring or including the quadrupole moment.  

The DMR data alone can not be used to constrain $\Omega_0$ over
range  $0.1 \leq \Omega_0 \leq 1$ in a statistically 
meaningful fashion for the open models considered here. It is, however,
reasonable to conclude that when the quadrupole moment is excluded from the
analysis, the $\Omega_0 \sim 0.4$ model CMB anisotropy spectral shape 
is most consistent with the DMR data, while the quadrupole-included analysis
favours $\Omega_0 \sim 0.7$ (for the open-bubble inflation model in the range
$0.1 \le \Omega_0 \leq 1$).  

Current cosmographic observations, in conjunction with current large-scale 
structure observations compared to the predictions of the DMR-normalized 
open-bubble inflation model derived here, favour $0.3 < \Omega_0 \lap 0.6$. The 
large allowed range is partially a consequence of the current uncertainty
in $\Omega_B$. This range is consistent with the value weakly favoured 
($\Omega_0 \sim 0.4$) by a quadrupole-excluded analysis of the DMR data alone. 
It might also be significant that mild bias is indicated both by the need to 
reconcile these larger values of $\Omega_0$ 
with what is determined from small-scale dynamical estimates, as well as to 
reconcile the smaller DMR-normalized $(\delta M/M)[ 8h^{-1}\ {\rm Mpc}]$
values (for this favoured range of $\Omega_0$) with the larger observed
galaxy number fluctuations (e.g., eq. [18]).  

In common with the low-density flat-$\Lambda$ CDM model, we have established 
that in the low-density open-bubble CDM model one may adjust the value of 
$\Omega_0$ to accommodate a large fraction of present observational 
constraints. For a broad class of these models, with adiabatic gaussian 
initial energy-density perturbations, this focuses attention on values of 
$\Omega_0$ that are larger than the range of values for $\Omega_B$ inferred 
from the observed light-element abundances in conjunction with standard 
nucleosynthesis theory. Whether this additional CDM is nonbaryonic, or is 
simply baryonic material that does not take part in standard nucleosynthesis,
remains a major outstanding puzzle for these models.

In conclusion, the open-bubble inflation model with $0.3 < \Omega_0 \lap
0.6$ is most consistent with current observations.

\bigskip

We acknowledge the efforts of those contributing to the $COBE$-DMR. $COBE$ is 
supported by the Office of Space Sciences of NASA Headquarters. We also 
acknowledge the advice and assistance of C. Baugh, S. Cole, J. Garriga, 
T. Kolatt, C. Park, L. Piccirillo, G. Rocha, G. Tucker, D. Weinberg, and 
K. Yamamoto. RS is supported in part by a PPARC grant and KBN grant 2P30401607.

\clearpage

\begin{center}
FIGURE CAPTIONS
\end{center}

\vspace{7mm}
\noindent Fig. 1.--Fractional differences, $\Delta C_\ell/C_\ell$, between 
the CMB spatial anisotropy multipole coefficients $C_\ell$ computed using the 
two Boltzmann transfer codes (and normalized to agree at $\ell = 9$). Heavy 
type is for the open-bubble inflation model spectrum accounting only for 
perturbations that are generated during the evolution inside the bubble 
(type [1] spectra above), and light type is for the open-bubble inflation 
model spectrum now also accounting for perturbations generated in the first 
epoch of inflation (type [2] spectra). Solid lines are for $\Omega_0 = 0.2$ 
and dashed lines are for 
$\Omega_0 = 0.5$. These are for $h = 0.6$ and $\Omega_B = 0.035$. Note that
$\Delta Q_{\rm rms-PS}/Q_{\rm rms-PS} \simeq 0.5 \Delta C_2/C_2$.

\vspace{3mm}
\noindent Fig. 2.--(a) CMB anisotropy multipole coefficients for the 
open-bubble inflation model, accounting only for fluctuations generated during 
the evolution inside the bubble (RP94, solid lines), and also accounting for 
fluctuations generated in the first epoch of inflation (BGT; YST, 
dotted lines --- these overlap the solid lines, except at the lowest 
$\Omega_0$ and smallest $\ell$), for $\Omega_0=$ 0.1, 0.2, 0.25, 0.3, 0.35, 
0.4, 0.45, 0.5, 0.6, 0.8, and 1.0, in ascending order. These are for 
$t_0 \simeq 12$ Gyr and $\Omega_B h^2 = 0.0125$. The coefficients are 
normalized relative to the $C_9$ amplitude, and different values of $\Omega_0$ 
are offset from each other to aid visualization. In (b) are the set of CMB
anisotropy spectra for the open-bubble inflation model, accounting only for
fluctuations generated during the evolution inside the bubble (RP94), with 
$\Omega_0 = 0.2$ and $\Omega_0 = 0.5$ for 
the three different pairs of values ($t_0$, $\Omega_B h^2$): ($\simeq 10.5$ 
Gyr, $= 0.0055$), ($\simeq 12$ Gyr, $= 0.0125$), and ($\simeq 13.5$ Gyr, 
$= 0.0205$). Spectra in the two sets are normalized to have the same $C_9$, 
and $\Omega_B h^2$ increases in ascending order on the right axis. 

\vspace{3mm} 
\noindent Fig. 3.--CMB spatial anisotropy multipole coefficients for the 
flat-space scale-invariant spectrum open model (W83). 
Conventions and parameter values are as in the caption of Fig. 2 (although
only one set of spectra are shown in Fig. 3a).

\vspace{3mm}
\noindent Fig. 4.--CMB spatial anisotropy multipole coefficients for the 
open-bubble inflation spectrum, also accounting for both fluctuations 
generated in the first epoch of inflation and that corresponding to a 
non-square-integrable basis function (YST, solid lines),
and ignoring both these fluctuations (RP94, dotted lines). 
They are, in ascending order, for $\Omega_0=$ 0.1 to 0.9 in steps of 0.1, 
with $h = 0.6$ and $\Omega_B = 0.035$, normalized relative to the $C_9$ 
amplitude, and different values of $\Omega_0$ are offset from each other to 
aid visualization.

\vspace{3mm}
\noindent Fig. 5.--CMB spatial anisotropy multipole coefficients, as a 
function of $\ell$, for the 
various spectra considered in this paper, at $\Omega_0 = 0.2$ and $0.5$ 
(vertically offset). Light solid and heavy solid lines show the open-bubble 
inflation cases accounting for (type [2] spectra above) 
and ignoring (type [1] spectra, at $\Omega_0 = 0.5$ these completely 
overlap the type [2] spectra) fluctuations 
generated in the first epoch of inflation. Dashed lines show the open-bubble 
inflation models, now also accounting for the contribution from the 
non-square-integrable basis function (type [3] spectra). Dotted lines show 
the flat-space scale-invariant spectrum open model spectra (type [4] spectra). 
All spectra are for $h = 0.6$ and $\Omega_B = 0.035$.

\vspace{3mm}
\noindent Fig. 6.--Likelihood functions $L(Q_{\rm rms-PS}, \Omega_0)$
(arbitrarily normalized to unity at the highest peak at $\Omega_0 \sim
0.4$) derived from a simultaneous analysis of the DMR 53 and 90 GHz 
ecliptic-frame data, ignoring the correction for faint high-latitude foreground
Galactic emission, and excluding the quadrupole moment from the 
analysis. These are for the $h = 0.6$, $\Omega_B = 0.035$ models. Panel
(a) is for the flat-space scale-invariant spectrum open model (W83), 
(b) is for the open-bubble inflation model accounting only for 
perturbations generated during the evolution inside the bubble (RP94), 
and (c) is for the open-bubble inflation model now also accounting for both 
the fluctuations generated in the first epoch of inflation and
those corresponding to a non-square-integrable basis function (YST).

\vspace{3mm}
\noindent Fig. 7.--Likelihood functions $L(Q_{\rm rms-PS}, \Omega_0)$
(arbitrarily normalized to unity at the highest peak near either
$\Omega_0 \sim 0.1$ or $0.7$), derived from a simultaneous analysis of the 
DMR 53 and 90 GHz galactic-frame data, accounting for the faint high-latitude
foreground Galactic emission correction, and including the quadrupole 
moment in the analysis. Conventions and parameter values are as 
for Fig. 6.

\vspace{3mm}
\noindent Fig. 8.--Ridge lines of the maximum likelihood $Q_{\rm rms-PS}$
value as a function of $\Omega_0$, for the open-bubble inflation model 
accounting only for fluctuations generated during the evolution inside 
the bubble (type [1] spectra), for the eight different DMR data sets 
considered here, and for $t_0 \simeq 12$ Gyr, $\Omega_B h^2 = 0.0125$.
Heavy lines correspond to the case when the quadrupole moment is 
excluded from the analysis, while light lines account for the quadrupole 
moment. These are for the ecliptic-frame sky maps, accounting for (dashed
lines) and ignoring (solid lines) the faint high-latitude foreground 
Galactic emission correction, and for the galactic-frame maps, accounting 
for (dot-dashed lines) and ignoring (dotted lines) this Galactic emission
correction. The general features of this figure are consistent with that
derived from the DMR two-year data (GRSB, Fig. 2).

\vspace{3mm}
\noindent Fig. 9.--Ridge lines of the maximum likelihood $Q_{\rm rms-PS}$
value as a function of $\Omega_0$, for the flat-space scale-invariant
spectrum open model (type [4] spectra), for the eight different DMR data sets, 
and for $t_0 \simeq 12$ Gyr, $\Omega_B h^2 = 0.0125$. 
Heavy lines correspond to the ecliptic-frame analyses, while light lines 
are from the galactic-frame analyses. These are for the cases ignoring the 
faint high-latitude foreground Galactic-emission correction, and either 
including (dotted lines) or excluding (solid lines) the quadrupole 
moment; and accounting for this Galactic emission correction, and either
including (dot-dashed lines) or excluding (dashed lines) the quadrupole
moment. The general features of this figure are roughly 
consistent with that derived from the DMR two-year data (Cay\'on et al. 1996,
Fig. 3). 

\vspace{3mm} 
\noindent Fig. 10.--Ridge lines of the maximum likelihood $Q_{\rm rms-PS}$
value as a function of $\Omega_0$, for the open-bubble inflation model now
also accounting for both the fluctuations generated in the first epoch 
of inflation (BGT; YST) and those from a 
non-square-integrable basis function (YST), for the eight
different DMR data sets considered here, and for $h = 0.6$, $\Omega_B = 0.035$.
Heavy lines correspond to the cases where the faint high-latitude foreground
Galactic emission correction is ignored, while light lines account for this
Galactic emission correction. These are from the ecliptic frame analyses,
accounting for (dotted lines) or ignoring (solid lines) the quadrupole 
moment; and from the galactic-frame analyses, accounting for (dot-dashed
lines) or ignoring (dashed lines) the quadrupole moment. The 
general features of this figure are consistent with that derived from the 
DMR two-year data (YB, Fig. 2). 

\vspace{3mm}
\noindent Fig. 11.--Ridge lines of the maximum likelihood $Q_{\rm rms-PS}$
value as a function of $\Omega_0$, for the two extreme DMR data sets, and two
different CMB anisotropy angular spectra, showing the effects of varying 
$t_0$ and $\Omega_B h^2$. Heavy lines are for $t_0 \simeq 13.5$ Gyr and 
$\Omega_B h^2 = 0.0205$, while light lines are for $t_0 \simeq 10.5$ Gyr
and $\Omega_B h^2 = 0.0055$. Two of the four pairs of lines are for the 
open-bubble inflation model accounting only for fluctuations generated 
during the evolution inside the bubble (type [1] spectra), either from 
the ecliptic-frame analysis without the faint high-latitude foreground Galactic
emission correction and ignoring the quadrupole moment in the analysis (solid 
lines), or from the galactic-frame analysis accounting for this Galactic 
emission correction and including the quadrupole moment in the analysis (dotted
lines). The other two of the four pairs of lines are for the flat-space
scale-invariant spectrum open model (type [4] spectra), either from the 
ecliptic-frame analysis without the faint high-latitude foreground Galactic 
emission correction and ignoring the quadrupole moment in the analysis (dashed 
lines), or from the galactic-frame analysis accounting for this Galactic
emission correction and including the quadrupole moment in the analysis 
(dot-dashed lines). Given the other uncertainties, the effects of varying 
$t_0$ and $\Omega_B h^2$ are fairly negligible. 

\vspace{3mm}
\noindent Fig. 12.--Ridge lines of the maximum likelihood $Q_{\rm rms-PS}$
value as a function of $\Omega_0$, for the two extreme DMR data sets, for
the four CMB anisotropy angular spectra models considered here, and for
$h = 0.6$, $\Omega_B = 0.035$. Heavy lines are from the ecliptic-frame
sky maps ignoring the faint high-latitude foreground Galactic emission 
correction and excluding the quadrupole moment from the analysis, while 
light lines are from the galactic-frame sky maps accounting for this
Galactic emission correction and including the quadrupole moment in the 
analysis. Solid, dotted, and dashed lines show the open-bubble inflation
cases, accounting only for the fluctuations generated during the evolution
inside the bubble (type [1] spectra, solid lines), also accounting for
the fluctuations generated in the first epoch of inflation
(type [2] spectra, dotted lines --- these overlap
the solid lines except for $\Omega_0 \lap 0.2$ and $\Omega_0 \sim 0.7$),
and finally also accounting for the fluctuations corresponding to the 
non-square-integrable basis function (type [3] spectra, dashed lines).
Dot-dashed lines correspond to the flat-space scale-invariant spectrum 
open model (type [4] spectra).

\vspace{3mm}
\noindent Fig. 13.--Conditional likelihood densities for $Q_{\rm rms-PS}$,
derived from $L(Q_{\rm rms-PS}, \Omega_0)$ (which are normalized to be 
unity at the peak, for each DMR data set, CMB anisotropy angular spectrum, and 
set of model-parameter values). Panel (a) is for the open-bubble inflation 
model accounting only for fluctuations generated during the evolution inside
the bubble (type [1] spectra), while panel (b) is for the flat-space
scale-invariant spectrum open model (type [4] spectra). The heavy lines are for
$\Omega_0 = 0.2$, while the light lines are for $\Omega_0 = 0.5$. Two of the
four pairs of lines in each panel correspond to the results from the analysis
of the galactic-frame maps accounting for the faint high-latitude foreground
Galactic emission correction and with the quadrupole moment 
included in the analysis,
either for $t_0 \simeq 10.5$ Gyr and $\Omega_B h^2 = 0.0055$ (dot-dashed 
lines), or for $t_0 \simeq 13.5$ Gyr and $\Omega_B h^2 = 0.0205$ (dashed lines).
The other two pairs of lines in each panel correspond to the results from the
analysis of the ecliptic-frame maps ignoring this Galactic emission correction
and with the quadrupole moment excluded from the analysis, either for 
$t_0 \simeq 10.5$ Gyr and $\Omega_B h^2 = 0.0055$ (dotted lines), or for 
$t_0 \simeq 13.5$ Gyr and $\Omega_B h^2 = 0.0205$ (solid lines). Given the
other uncertainties, the effects of varying $t_0$ and $\Omega_B h^2$ are 
fairly negligible.

\vspace{3mm}
\noindent Fig. 14.--Conditional likelihood densities for $Q_{\rm rms-PS}$
normalized as in the caption for Fig. 13. Panel (a) is from the analysis 
of the ecliptic-frame maps ignoring the faint high-latitude foreground
Galactic emission correction and excluding the quadrupole moment from the
analysis, while panel (b) is from the analysis of the galactic-frame 
maps accounting for this Galactic emission correction and including the 
quadrupole moment in the analysis. These are for $h = 0.6$ and
$\Omega_B = 0.035$. The heavy lines are for $\Omega_0 = 0.2$ and the 
light lines are for $\Omega_0 = 0.5$. There are eight lines (four pairs)
in each panel, although in each panel two pairs almost identically
overlap. Solid, dotted, and dashed lines show the open-bubble inflation
cases, accounting only for the fluctuations generated during the evolution 
inside the bubble (type [1] spectra, solid lines), also accounting for the 
fluctuations generated in the first epoch of inflation (type [2] spectra, 
dotted lines --- these
almost identically overlap the solid lines), and finally also accounting
for the fluctuations corresponding to the non-square-integrable basis 
function (type [3] spectra, dashed lines). Dot-dashed lines correspond
to the flat-space scale-invariant spectrum open model (type [4] spectra).

\vspace{3mm}
\noindent Fig. 15.--Projected likelihood densities for $\Omega_0$ 
derived from $L(Q_{\rm rms-PS}, \Omega_0)$ (normalized as in the 
caption of Fig. 13). Panel (a) is for the open-bubble inflation model
accounting only for the fluctuations generated during the evolution 
inside the bubble (type [1] spectra), and panel (b) is for the 
flat-space scale-invariant spectrum open model (type [4] spectra). Two of the
curves in each panel correspond to the results from the analysis of the 
galactic-frame maps accounting for the faint high-latitude foreground
Galactic emission correction and with the quadrupole moment 
included in the analysis,
for $t_0 \simeq 10.5$ Gyr and $\Omega_B h^2 = 0.0055$ (dot-dashed lines)
and for $t_0 \simeq 13.5$ Gyr and $\Omega_B h^2 = 0.0205$ (dashed lines).
The other two curves in each panel are from the analysis of the ecliptic-frame
maps ignoring the Galactic emission correction and excluding the 
quadrupole moment from the analysis, for $t_0 \simeq 10.5$ Gyr and 
$\Omega_B h^2 = 0.0055$ (dotted lines) and for $t_0 \simeq 13.5$ Gyr
and $\Omega_B h^2 = 0.0205$ (solid lines).

\vspace{3mm}
\noindent Fig. 16.--Projected likelihood densities for $\Omega_0$ derived
from $L(Q_{\rm rms-PS}, \Omega_0)$ (normalized as in the caption of Fig. 13).
Panel (a) is from the analysis of the ecliptic-frame sky maps ignoring the
faint high-latitude foreground Galactic emission correction and excluding
the quadrupole moment from the analysis. Panel (b) is from the analysis of
the galactic-frame sky maps accounting for this Galactic emission correction
and including the quadrupole moment in the analysis. There are four curves
in each panel, although in each panel two of them almost overlap. Solid,
dotted, and dashed lines show the open-bubble inflation cases, accounting 
only for the fluctuations generated during the evolution inside the bubble
(type [1] spectra, solid lines), also accounting for the fluctuations
generated in the first epoch of spatially-flat inflation (type [2] spectra, 
dotted lines --- these almost exactly overlap the solid lines), and finally 
also accounting for the fluctuations corresponding to the 
non-square-integrable basis function (type [3] spectra, dashed lines). 
Dot-dashed lines correspond to the flat-space scale-invariant 
spectrum open model (type [4] spectra). These are for $h = 0.6$ and $\Omega_B =
0.035$.

\vspace{3mm}
\noindent Fig. 17.--Marginal likelihood densities [$\propto \int dQ_{\rm rms-PS}
\, L(Q_{\rm rms-PS}, \Omega_0)$] for $\Omega_0$, normalized to unity at the
peak, for the open-bubble inflation model accounting only for fluctuations
generated during the evolution inside the bubble (RP94),
for the eight different DMR data sets, and for $t_0 \simeq 12$ Gyr, 
$\Omega_B h^2 = 0.0125$. Panel (a) is from the ecliptic-frame analyses,
and panel (b) is from the galactic-frame analyses. Two of the four lines in
each panel are from the analysis without the faint high-latitude foreground
Galactic emission correction, either accounting for (dot-dashed lines) or 
ignoring (solid lines) the quadrupole moment. The other two lines in 
each panel are from the analysis with this Galactic emission correction,
either accounting for (dotted lines) or ignoring (dashed lines) the quadrupole 
moment. 

\vspace{3mm}
\noindent Fig. 18.--Marginal likelihood densities for $\Omega_0$, for
the flat-space scale-invariant spectrum open model (W83). Conventions
and parameter values are as in the caption of Fig. 17.

\vspace{3mm}
\noindent Fig. 19.--Marginal likelihood densities for $\Omega_0$, for the
open-bubble inflation model now also accounting for both the fluctuations
generated in the first spatially-flat epoch of inflation and those that 
correspond to the non-square-integrable basis function (YST), computed for 
$h = 0.6$ and $\Omega_B = 0.035$. Conventions are as in the caption of Fig. 17.

\vspace{3mm}
\noindent Fig. 20.--Marginal likelihood densities for $\Omega_0$ (normalized
as in the caption of Fig. 17). Panel (a) is for the open-bubble inflation
model accounting only for the fluctuations generated during the evolution 
inside the bubble (RP94), while panel (b) is for the 
flat-space scale-invariant spectrum open model (W83). Two of the lines
in each panel are the results from the analysis of the galactic-frame data
sets accounting for the faint high-latitude foreground Galactic emission 
correction and with the quadrupole moment included in the analysis, for 
$t_0 \simeq 10.5$ Gyr and $\Omega_B h^2 = 0.0055$ (dot-dashed lines),
and for $t_0 \simeq 13.5$ Gyr and $\Omega_B h^2 = 0.0205$ (dashed lines).
The other two lines in each panel are the results from the analysis of
the ecliptic-frame data sets ignoring this Galactic emission correction 
and with the quadrupole moment excluded from the analysis, for $t_0 \simeq
10.5$ Gyr and $\Omega_B h^2 = 0.0055$ (dotted lines), and for $t_0 \simeq
13.5$ Gyr and $\Omega_B h^2 = 0.0205$ (solid lines).

\vspace{3mm}
\noindent Fig. 21.--Marginal likelihood densities for $\Omega_0$ (normalized 
as in the caption of Fig. 17), computed for $h = 0.6$ and $\Omega_B = 0.035$. 
Panel (a) is from the analysis of the 
ecliptic-frame sky maps ignoring the faint high-latitude foreground 
Galactic emission correction and excluding the quadrupole moment from
the analysis. Panel (b) is from the analysis of the galactic-frame sky maps
accounting for this Galactic emission correction and including the 
quadrupole moment in the analysis. There are four lines in each panel, although
in each panel two of the lines almost overlap. Solid, dotted, and dashed 
curves are the open-bubble inflation cases, accounting only for the 
fluctuations generated during the evolution inside the bubble (RP94, solid 
lines), also accounting for the fluctuations generated in the first epoch of 
spatially-flat inflation (BGT; YST, dotted lines --- these almost identically 
overlap the solid lines), and finally also accounting for the fluctuations 
corresponding to the non-square-integrable basis function (YST, dashed lines). 
Dot-dashed curves correspond to the flat-space scale-invariant spectrum open 
model (W83). 

\vspace{3mm}
\noindent Fig. 22.--Fractional differences, $\Delta P(k)/P(k)$, as a function
of wavenumber $k$, between the energy-density perturbation power spectra $P(k)$ 
computed using the two independent numerical integration codes (and normalized 
to give the same $Q_{\rm rms-PS}$). The heavy curves are for the open-bubble 
inflation model spectrum accounting only for fluctuations that are generated 
during the evolution inside the bubble (type [1] spectra), and the 
light curves are for the open-bubble inflation model spectrum now also 
accounting for fluctuations generated in the first epoch of 
inflation (type [2] spectra). These are for $\Omega_0 = 0.2$ (solid lines) and 
$\Omega_0 = 0.5$ (dashed lines), with $h = 0.6$ and $\Omega_B = 0.035$. 

\vspace{3mm}
\noindent Fig. 23.--Fractional energy-density perturbation power spectra
$P(k)$ as a function of wavenumber $k$. These are normalized to the 
mean of the extreme upper and lower 2-$\sigma$ $Q_{\rm rms-PS}$ values
(as discussed in \S 3.3).
Panels (a)--(d) correspond to the four different sets of ($t_0$, $\Omega_B 
h^2$) of Tables 9--12, and each panel shows power spectra
for three different models at six values of $\Omega_0$. Solid lines show
the open-bubble inflation model $P(k)$ accounting only for fluctuations
generated during the evolution inside the bubble (RP95); dotted lines are for 
the open-bubble inflation model now also accounting for fluctuations generated 
in the first epoch of inflation (BGT; YST); and, dashed lines are for the 
flat-space scale-invariant spectrum open model (W83). Starting near the
center of the lower horizontal axis, and moving counterclockwise, the spectra
shown correspond to $\Omega_0=$ 0.1, 0.2, 0.3, 0.45, 0.6, and 1. Note that at
$\Omega_0 = 1$ all three model spectra are identical and so overlap; also note 
that at a given $\Omega_0$ the open-bubble inflation model $P(k)$ accounting 
for the fluctuations generated in the first epoch of inflation (BGT; YST, 
dotted lines) essentially overlap those where this source of fluctuations is 
ignored (RP95, solid lines). Panel (a) corresponds to $t_0 \simeq 10.5$ 
Gyr and $\Omega_B h^2 = 0.0055$, (b) to $t_0 \simeq 12$ Gyr and $\Omega_B h^2 
= 0.0125$, (c) to $t_0 \simeq 13.5$ Gyr and $\Omega_B h^2 = 0.205$, and (d) 
to $t_0 \simeq 12$ Gyr and $\Omega_B h^2 = 0.007$ (normalized using the 
results of the DMR analysis of the $t_0 \simeq 10.5$ Gyr, $\Omega_B h^2 =
0.0055$ models). Panel (e) shows the three $h = 0.6$, $\Omega_B = 0.035$ 
open-bubble inflation spectra of Table 13 at five different values of 
$\Omega_0$. The spectra are for the open-bubble inflation model accounting 
only for fluctuations generated during the evolution inside the bubble (RP95, 
solid lines), also accounting for fluctuations generated in the first epoch of 
inflation (BGT; YST, dotted lines), and also accounting for the contribution 
from the non-square-integrable basis function (YST, dashed
lines). Starting near the center of the lower horizontal axis and moving
counterclockwise, the models correspond to $\Omega_0=$ 0.1, 0.2, 0.3, 0.5,
and 0.9. Note that at a given $\Omega_0$ the three spectra essentially overlap,
especially for observationally-viable values of $\Omega_0 \gap 0.3$. The solid 
triangles represent the redshift-space da Costa et al. (1994) SSRS2 + CfA2 
($130h^{-1}$ Mpc depth) optical galaxies data (and were very kindly provided 
to us by C. Park). The solid squares represent the [$P(k) = 8000
(h^{-1}\ {\rm Mpc})^3$ weighting] redshift-space results of the Tadros
\&\ Efstathiou (1995) analysis of the $IRAS$ QDOT and 1.2 Jy infrared galaxy 
data. The hollow pentagons represent the real-space results of the Baugh
\&\ Efstathiou (1993) analysis of the APM optical galaxy data (and were very
kindly provided to us by C. Baugh). It should be noted that the plotted model
mass (not galaxy) power spectra do not account for any bias of galaxies 
with respect to mass. They also do not account for nonlinear or 
redshift-space-distortion (when relevant) corrections nor for the survey 
window functions. It should also be noted that the observational data error 
bars are determined under the assumption of a specific cosmological model
and a specific evolution scenario, i.e., they do not necessarily account 
for these additional sources of uncertainty (e.g., Gazta\~naga 1995). We
emphasize that, because of the different assumptions, the different observed 
galaxy power spectra shown on the plots are defined somewhat differently and 
so cannot be directly quantitatively compared to each other.

\vspace{3mm}
\noindent Fig. 24.--CMB anisotropy bandtemperature predictions and 
observational results, as a function of multipole $\ell$, to $\ell = 1000$. 
The four pairs of wavy curves (in different linestyles) demarcating the
boundaries of the four partially overlapping wavy hatched regions (hatched 
with straight lines in different linestyles) in panel (a) are DMR-normalized
open-bubble inflation model (RP94) predictions for what would be seen by a 
series of ideal, Kronecker-delta window-function, experiments (see Ratra et 
al. 1995 for details). Panel (b) shows DMR-normalized CMB anisotropy spectra 
with the same cosmological parameters for the flat-space scale-invariant 
spectrum open model (W83). The model-parameter values 
are: $\Omega_0 = 0.3$, $h = 0.7$, $\Omega_B h^2 = 0.0075$, $t_0 = 11.3$ Gyr 
(dot-dashed lines); $\Omega_0 = 0.4$, $h = 0.65$, $\Omega_B h^2 = 0.0125$,
$t_0 = 11.7$ Gyr (solid lines); $\Omega_0 = 0.5$, $h = 0.55$, $\Omega_B h^2
= 0.0175$, $t_0 = 13.4$ Gyr (dashed lines); and, $\Omega_0 = 1$, $h = 0.5$,
$\Omega_B h^2 = 0.0125$, $t_0 = 13.0$ Gyr (dotted lines) --- for more details 
on these models see Ratra et al. (1995). For each pair of model-prediction
demarcation curves, the lower one is normalized to the lower 1-$\sigma$ 
$Q_{\rm rms-PS}$
value determined from the analysis of the galactic-coordinate maps accounting
for the high-latitude Galactic emission correction and including the $\ell =
2$ moment in the analysis, and the upper one is normalized to the upper 
1-$\sigma$ $Q_{\rm rms-PS}$ value determined from the analysis of the 
ecliptic-coordinate maps ignoring the Galactic emission correction and 
excluding the $\ell = 2$ moment from the analysis. Amongst the open-bubble
inflation models of panel (a), the $\Omega_0 = 0.4$
model is close to what is favoured by the analysis of Table 10, and the 
$\Omega_0 = 0.5$ model is close to that preferred from the analysis of Table 
11. The $\Omega_0 = 0.3$ model is on the edge of the allowed region from the
analysis of Table 12, and the $\Omega_0 = 1$ fiducial CDM model is incompatible
with cosmographic and large-scale structure observations. A large fraction of 
the smaller-scale observational data in these plots are tabulated in Ratra et 
al. (1995) and Ratra \&\ Sugiyama (1995). Note that, as discussed in these 
papers, some of the data points are from reanalyses of the observational data.
There are 69 detections and 22 2-$\sigma$ upper limits shown. Since most of the 
smaller-scale data points are derived assuming a flat bandpower CMB anisotropy
angular spectrum, which is more accurate for narrower (in $\ell$) window
functions, we have shown the observational results from the narrowest windows
available. The data shown are from the DMR galactic frame maps ignoring
the Galactic emission correction (G\'orski 1996, open octagons with $\ell 
\leq 20$); from FIRS (Ganga et al. 1994, as analyzed by Bond 1995, solid 
pentagon); Tenerife (Hancock et al. 1996a, open five-point star); Bartol
(Piccirillo et al. 1996, solid diamond, note that atmospheric contamination
may be an issue); SK93, individual-chop SK94 Ka and Q, and individual-chop 
SK95 cap and ring (Netterfield et al. 1996, open squares); SP94 Ka and Q 
(Gundersen et al. 1995, the points plotted here are from the flat bandpower 
analysis of Ganga et al. 1996a, solid circles); BAM 2-beam (Tucker et al. 
1996, at $\ell_{\rm eff} = 58.2$ with $\ell_{e^{-0.5}}$ spanning 16 to 92, 
and accounting for the $20\%$ calibration uncertainty, open circle);
Python-G, -L, and -S (e.g., Platt et al. 1996, open six-point stars); ARGO 
(e.g., Masi et al. 1996, both the Hercules and Aries+Taurus scans are shown 
--- note that the Aries+Taurus scan has a larger calibration uncertainty of 
$10\%$, solid squares); MAX3, individual-channel MAX4, and MAX5 (e.g., Tanaka 
et al. 1996, including the MAX5 MUP 2-$\sigma$ upper limit $\delta T_\ell < 
35\ \mu$K at $\ell_{\rm eff} = 139$, Lim et al. 1996, open hexagons); MSAM92 
and MSAM94 (e.g., Inman et al. 1996, open diamonds); WDH1--3 and WDI, II 
(e.g., Griffin et al. 1996, open pentagons); and CAT (Scott et al. 1996 --- 
CAT1 at $\ell_{\rm eff} = 396$ with $\ell_{e^{-0.5}}$ spanning 351 to 471, and
CAT2 at $\ell_{\rm eff} = 608$ with $\ell_{e^{-0.5}}$ spanning 565 to 710,
both accounting for calibration uncertainty of $5\%$, solid hexagons).
Detections have vertical 1-$\sigma$ error bars. Solid inverted triangles 
inserted inside the appropriate symbols correspond to nondetections, and are 
placed at the upper 2-$\sigma$ limits. Vertical error bars are not shown for 
non-detections. As discussed in Ratra et al. (1995), all $\delta T_\ell$
(vertical) error bars
also account for the calibration uncertainty (but in an approximate manner,
except for the SP94 Ka and Q results from Ganga et al. 1996a --- see Ganga
et al. 1996a for a discussion of this issue). The observational data points
are placed at the $\ell$-value at which the corresponding window function
is most sensitive (this ignores the fact that the sensitivity of the 
experiment is also dependent on the assumed form of the sky-anisotropy signal,
and so gives a somewhat misleading impression of the multipoles to which the
experiment is sensitive --- see Ganga et al. 1996a for a discussion of this 
issue). Excluding the DMR points at $\ell \leq 20$, the horizontal lines on the 
observational data points represent the $\ell$-space width of the 
corresponding window function (again ignoring the form of the sky-anisotropy 
signal). Note that from an analysis of a large fraction of the data 
(corresponding to detections of CMB anisotropy) shown in these figures,
GRS (Figs. 5 and 6) conclude that all the models shown in panel (a),
including the fiducial CDM one, are consistent with the CMB anisotropy 
data.


\clearpage

\begin{references}

\ref Abbott, L.F., \&\ Schaefer, R.K.;1986;ApJ;308;546

\ref Bahcall, N.A., \&\ Oh, S.P.;1996;ApJ;462;L49

\rep Banday, A.J., et al.;1996a;ApJ, submitted

\rep Banday, A.J., et al.;1996b;ApJ, in press

\ref Banks, T., et al.;1995;Phys. Rev. D;52;3548

\ref Bardeen, J.M., Bond, J.R., Kaiser, N., \&\ Szalay, A.S.;1986;ApJ;304;15
(BBKS)

\ref Baugh, C.M.;1996;MNRAS;280;267

\ref Baugh, C.M., \&\ Efstathiou, G.;1993;MNRAS;265;145 

\ref Baum, W.A., et al.;1995;AJ;110;2537

\ref Bennett, C.L., et al.;1994;ApJ;436;423

\ref Bennett, C.L., et al.;1996;ApJ;464;L1

\ref Bertschinger, E., et al.;1990;ApJ;364;370

\ref Bond, J.R.;1995;Phys. Rev. Lett.;74;4369

\rep Bond, J.R.;1996;in Cosmology and Large Scale Structure, ed. R. Schaeffer,
J. Silk, \&\ J. Zinn-Justin (Dordrecht: Elsevier Science Publishers), in press

\rep Branch, D., Fisher, A., Baron, E., \&\ Nugent, P.;1996;Nature, submitted

\ref Bucher, M., Goldhaber, A.S., \&\ Turok, N.;1995;Phys. Rev. D;52;3314 (BGT)

\ref Bucher, M., \&\ Turok, N.;1995;Phys. Rev. D;52;5538 (BT)

\rep Bunn, E.F., Liddle, A.R., \&\ White, M.;1996;astro-ph/9607038

\rep Bunn, E.F., \&\ White, M.;1996;astro-ph/9607060 (BW)

\ref Buote, D.A., \&\ Canizares, C.R.;1996;ApJ;457;565

\rep Burles, S., \&\ Tytler, D.;1996;Science, submitted

\rep Cardall, C.Y., \&\ Fuller, G.M.;1996;ApJ, in press

\ref Carlberg, R.G., et al.;1996;ApJ;462;32

\ref Carswell, R.F., et al.;1994;MNRAS;268;L1

\ref Carswell, R.F., et al.;1996;MNRAS;278;506

\ref Cay\'on, L., et al.;1996;MNRAS;279;1095

\ref Cen, R., \&\ Ostriker, J.P.;1993;ApJ;414;407

\ref Chaboyer, B., Demarque, P., Kernan, P.J., \&\ Krauss, 
L.M.;1996;Science;271;957 

\rep Cohn, J.D.;1996;LBL preprint LBNL-38560

\ref Cole, S., Fisher, K.B., \&\ Weinberg, D.H.;1995;MNRAS;275;515

\ref Copi, C.J., Schramm, D.N., \&\ Turner, M.S.;1995;Phys. Rev. Lett.;75;3981

\ref Cowie, L.L., Hu, E.M., \&\ Songaila, A.;1995;Nature;377;603

\ref Croft, R.A.C., \&\ Efstathiou, G.;1994;MNRAS;268;L23

\ref da Costa, L.N., et al.;1994;ApJ;437;L1

\ref Dar, A.;1995;ApJ;449;550

\ref David, L.P., Jones, C., \&\ Forman, W.;1995;ApJ;445;578

\rep Davis, M., Nusser, A., \&\ Willick, J.A.;1996;ApJ, submitted

\ref Djorgovski, S.G., Pahre, M.A., Bechtold, J., \&\ Elston, 
R.;1996;Nature;382;234

\ref Driver, S.P, Windhorst, R.A., Phillipps, S., \&\ Bristow, 
P.D.;1996;ApJ;461;525

\rep Eke, V.R., Cole, S., \&\ Frenk, C.S.;1996;MNRAS, submitted (ECF)

\ref Elbaz, D., Arnaud, M., B\"ohringer, H.;1995;A{\&}A;293;337

\rep Ellingson, E., Yee, H.K.C., Bechtold, J., \&\ Elston, R.;1996;ApJ, in 
press

\rep Fields, B.D., Kainulainen, K., Olive, K.A., \&\ Thomas, D.;1996;New
Astron., in press (FKOT)

\ref Fields, B.D., \&\ Olive, K.A.;1996;Phys. Lett. B;368;103

\ref Fischler, W., Ratra, B., \&\ Susskind, L.;1985;Nucl. Phys. B;259;730

\ref Fontana, A., et al.;1996;MNRAS;279;L27

\ref Francis, P.J., et al.;1996;ApJ;457;490

\ref Ganga, K., Page, L., Cheng, E., \&\ Meyer, S.;1994;ApJ;432;L15

\rep Ganga, K., Ratra, B., Gundersen, J.O., \&\ Sugiyama, N.;1996a;MIT preprint
MIT-CTP-2510

\ref Ganga, K., Ratra, B., \&\ Sugiyama, N.;1996b;ApJ;461;L61 (GRS)

\rep Garc\'{\i}a-Bellido, J.;1996;Phys. Rev. D, in press

\rep Garriga, J.;1996;Phys. Rev. D, in press

\ref Gazta\~naga, E.;1995;ApJ;454;561

\rep Giavalisco, M., Steidel, C.C., \&\ Macchetto, F.D.;1996; ApJ, in press

\ref Glazebrook, K., Peacock, J.A., Miller, L., \&\ Collins, C.A.;1995;MNRAS;
275;169

\ref G\'orski, K.M.;1994;ApJ;430;L85

\rep G\'orski, K.M.;1996; in preparation

\ref G\'orski, K.M., et al.;1996;ApJ;464;L11 (G96)

\ref G\'orski, K.M., et al.;1994;ApJ;430;L89

\ref G\'orski, K.M., Ratra, B., Sugiyama, N., \&\ Banday, A.J.;1995;ApJ; 
444;L65 (GRSB)

\ref Gott III, J.R.;1982;Nature;295;304

\rep Griffin, G.S., et al.;1996;in preparation

\ref Gouda, N., Sugiyama, N., \&\ Sasaki, M.;1991;Prog. Theo. Phys.;85;1023

\ref Gundersen, J.O., et al.;1995;ApJ;443;L57

\ref Guth, A.;1981;Phys. Rev. D;23;347

\ref Guth, A.H., \&\ Weinberg, E.J.;1983;Nucl. Phys. B;212;321

\rep Gwyn, S.D.J., \&\ Hartwick, F.D.A.;1996;ApJ, in press

\ref Hamazaki, T., Sasaki, M., Tanaka, T., \&\ Yamamoto, K.;1996;Phys. Rev.
D;53;2045

\rep Hancock, S., et al.;1996a;MNRAS, submitted

\rep Hancock, S., Rocha, G., Lasenby, A.N., \&\ Guti\'errez, C.M.;1996b;MNRAS,
submitted

\ref Harrison, E.R.;1970;Phys. Rev. D;1;2726

\ref Haslam, C.G.T., et al.;1981;A{\&}A;100;209

\ref Hu, W., \&\ Sugiyama, N.;1994;Phys. Rev. D;50;627

\ref Hu, W., \&\ Sugiyama, N.;1995;Phys. Rev. D;51;2599

\ref Im, M., Griffiths, R.E., Ratnatunga, K.J., \&\ Sarajedini, 
V.L.;1996;ApJ;461;L79

\rep Inman, C.A., et al.;1996;ApJ, submitted

\rep Jimenez, R., et al.;1996;MNRAS, in press

\ref Kamionkowski, M., Ratra, B., Spergel, D.N., \&\ Sugiyama, N.;1994;ApJ;
434;L1

\ref Kamionkowski, M., \&\ Spergel, D.N.;1994;ApJ;432;7

\ref Kazanas, D.;1980;ApJ;241;L59

\ref Kennicutt Jr., R.C., Freedman, W.L., \&\ Mould, J.R.;1995;AJ;110;1476

\rep Kitayama, T, \&\ Suto, Y.;1996;ApJ, in press

\ref Kochanek, C.S.;1996;ApJ;466;638

\ref Kogut, A., et al.;1996a;ApJ;464;L5

\rep Kogut, A., et al.;1996b;ApJ, in press

\rep Kolatt, T., \&\ Dekel, A.;1995;ApJ, submitted

\ref Liddle, A.R., Lyth, D.H., Roberts, D., \&\ Viana, 
P.T.P.;1996a;MNRAS;278;644 (LLRV)

\rep Liddle, A.R., Lyth, D.H., Viana, P.T.P, \&\ White, M.;1996b;MNRAS, in press

\ref Lilly, S.J., et al.;1995;ApJ;455;108

\rep Lim, M.A., et al.;1996;ApJ, in press

\ref Linde, A., \&\ Mezhlumian, A.;1995;Phys. Rev. D;52;6789

\rep Loveday, J., Efstathiou, G., Maddox, S.J., \&\ Peterson, B.A.;1996;ApJ,
in press

\ref Lu, L., Sargent, W.L.W., Womble, D.S., \&\ Barlow, T.A.;1996;ApJ;457;L1

\ref Luppino, G.A., \&\ Gioia, I.M.;1995;ApJ;445;L77

\ref Lyth, D.H., \&\ Woszczyna, A.;1995;Phys. Rev. D;52;3338

\ref Ma, C.-P., \&\ Bertschinger, E.;1994;ApJ;434;L5

\rep Maddox, S.J., Efstathiou, G., \&\ Sutherland, W.J.;1996;MNRAS, submitted

\ref Markevitch, M., et al.;1996;ApJ;456;437

\ref Masi, S., et al.;1996;ApJ;463;L47

\rep Moscardini, L., et al.;1995;MNRAS, submitted

\rep Netterfield, C.B., et al.;1996;ApJ, in press

\ref Nusser, A., \&\ Davis, M.;1994;ApJ;421;L1

\ref Ostriker, J.P., \&\ Cen, R.;1996;ApJ;464;27

\ref Ostriker, J.P., \&\ Steinhardt, P.J.;1995;Nature;377;600

\ref Pascarelle, S.M., et al.;1996;ApJ;456;L21

\ref Peacock, J.A., \&\ Dodds, S.J.;1994;MNRAS;267;1020 

\ref Peebles, P.J.E.;1984;ApJ;284;439

\rep Peebles, P.J.E.;1993;Principles of Physical Cosmology (Princeton: 
Princeton University Press)

\ref Peebles, P.J.E., \&\ Yu, J.T.;1970;ApJ;162;815

\rep Pell\'o, R., et al.;1996;A{\&}A, in press

\rep Perlmutter, S., et al.;1996; in Proceedings of NATO ASI on Thermonuclear
Supernovae, ed. R. Canal, P. Ruiz-Lapuente, \&\ J. Isern (Dordrecht: Kluwer),
in press

\rep Piccirillo, L., et al.;1996;ApJ, submitted

\rep Platt, S.R., et al.;1996;ApJ, submitted

\rep Ratcliffe, A., et al.;1996;Nature, submitted

\ref Ratra, B.;1991;Phys. Lett. B;260;21

\rep Ratra, B., Banday, A.J., G\'orski, K.M., \&\ Sugiyama, N.;1995;Princeton 
preprint PUPT-1558

\ref Ratra, B., \&\ Peebles, P.J.E.;1994;ApJ;432;L5 (RP94)

\ref Ratra, B., \&\ Peebles, P.J.E.;1995;Phys. Rev. D;52;1837 (RP95)

\ref Ratra, B., \&\ Quillen, A.;1992;MNRAS;259;738

\rep Ratra, B., \&\ Sugiyama, N.;1995;Princeton preprint PUPT-1559

\rep Reach, W.T., Franz, B.A., Kelsall, T., \&\ Weiland, J.L.;1995;in
Unveiling the Cosmic Infrared Background, ed. E. Dwek (New York: AIP), 37

\ref Renzini, A., et al.;1996;ApJ;465;L23

\rep Riess, A.G., Press, W.H., \&\ Kirshner, R.P.;1996;ApJ, submitted

\ref Rugers, M., \&\ Hogan, C.J.;1996a;ApJ;459;L1

\ref Rugers, M., \&\ Hogan, C.J.;1996b;AJ;111;2135

\ref Ruiz-Lapuente, P.;1996;ApJ;465;L83

\rep Salaris, M., Degl'Innocenti, S., \&\ Weiss, A.;1996;ApJ, submitted

\ref Sandage, A., et al.;1996;ApJ;460;L15

\rep Sarkar, S.;1996;Rep. Prog. Phys., submitted

\rep Sasaki, M., \&\ Tanaka, T.;1996;Osaka preprint OU-TAP-35

\ref Sato, K.;1981a;Phys. Lett. B;99;66

\ref Sato, K.;1981b;MNRAS;195;467

\ref Schaefer, B.E.;1996;ApJ;460;L19

\ref Scott, P.F., et al.;1996;ApJ;461;L1

\ref Shaya, E.J., Peebles, P.J.E., \&\ Tully, R.B.;1995;ApJ;454;15

\ref Smoot, G., et al.;1992;ApJ;396;L1

\ref Songaila, A., Cowie, L.L., Hogan, C.J., \&\ Rugers, M.;1994;Nature;368;599

\ref Steidel, C.C., et al.;1996;ApJ;462;L17

\ref Stompor, R.;1994;A{\&}A;287;693

\rep Stompor, R.;1996;in Microwave Background Anisotropies, ed. F.R. 
Bouchet (Dordrecht: Elsevier Science Publishers), to be published

\ref Stompor, R., G\'orski, K.M., \&\ Banday, A.J.;1995;MNRAS;277;1225

\ref Sugiyama, N.;1995;ApJS;100;281

\ref Sugiyama, N., \&\ Gouda, N.;1992;Prog. Theo. Phys.;88;803

\ref Sugiyama, N., \&\ Silk, J.;1994;Phys. Rev. Lett.;73;509

\ref Tadros, H., \&\ Efstathiou, G.;1995;MNRAS;276;L45

\rep Tanaka, S.T., et al.;1996;ApJ, in press

\ref Tanvir, N.R., Shanks, T., Ferguson, H.C., \&\ Robinson, 
D.R.T.;1995;Nature;377;27

\ref Torres, L.F.B., \&\ Waga, I.;1996;MNRAS;279;712

\rep Tucker, G.S., Gush, H., Halpern, M., \&\ Towlson, W.;1996;ApJ, submitted

\ref Tytler, D., Fan, X.-M., \&\ Burles, S.;1996;Nature;381;207

\ref van den Bergh, S.;1995;Science;270;1942

\ref Viana, P.T.P., \&\ Liddle, A.R.;1996;MNRAS;281;323 (VL)

\ref Vogt, N.P., et al.;1996;ApJ;465;L15

\rep Wampler, E.J., et al.;1996;A{\&}A, in press

\ref White, D.A., \&\ Fabian, A.C.;1995;MNRAS;273;72

\ref White, S.D.M., Navarro, J.F., Evrard, A.E., \&\ Frenk, 
C.S.;1993;Nature;366;429

\rep Williams, L.L.R., \&\ Lewis, G.F.;1996;MNRAS, submitted

\ref Wilson, M.L.;1983;ApJ;273;2 (W83)

\ref Wolfe, A.M.;1993;ANYAS;688;281

\ref Wright, E.L., et al.;1994;ApJ;420;1

\ref Wright, E.L., et al.;1996;ApJ;464;L21

\ref Yamamoto, K., \&\ Bunn, E.F.;1996;ApJ;464;8 (YB)

\ref Yamamoto, K., Sasaki, M., \&\ Tanaka, T.;1995;ApJ;455;412 (YST)

\rep Yamamoto, K., Sasaki, M., \&\ Tanaka, T.;1996; Hiroshima preprint HUPD 9604

\ref Yee, H.K.C., et al.;1996;AJ;111;1783

\rep Zaroubi, S., Dekel, A., Hoffman, Y., \&\ Kolatt, T.;1996;ApJ, submitted

\ref Zel'dovich, Ya. B.;1972;MNRAS;160;1P

\end{references}
\end{document}